\documentclass[twocolumn, twocolappendix]{aastex7}

\usepackage{gensymb}
\usepackage{soul}
\usepackage{graphicx}
\usepackage{tablefootnote}
\usepackage{verbatim}
\usepackage{xcolor}
\usepackage{array,multirow}
\usepackage{xspace} 

\usepackage{placeins}   
\usepackage{dblfloatfix} 

\shorttitle{Host Galaxies of AGNs with Direct Black Hole Mass Measurements}
\shortauthors{Bennert et al.}

\newcommand{\mbh}{$M_{\rm BH}$}

\newcommand\sigstar{$\sigma_\star$}

\newcommand\f{$f$}

\newcommand{\s}{{\rm s}}

\newcommand{\kms}{${\rm km\,s}^{-1}$}
\newcommand{\perMpc}{${\rm Mpc}^{-1}$}

\newcommand{\mdyn}{$M_{\rm sph, dyn}$}

\newcommand{\msun}{$M_{\odot}$}
\newcommand{\lsun}{$L_{\odot}$}

\begin{document}

\title{The Host Galaxies of Active Galactic Nuclei with Direct Black Hole Mass Measurements\footnote{Based on observations made with the NASA/ESA Hubble Space Telescope,
obtained at the Space Telescope Science Institute, which is operated by the
Association of Universities for Research in Astronomy, Inc., under NASA
contract NAS5-26555.
These observations are associated with programs
\# 17103 and 17063.
}
}
\author[0000-0003-2064-0518]{Vardha~N.~Bennert}
\affiliation{Physics Department, California Polytechnic State
  University, San Luis Obispo, CA 93407, USA}
\email{vbennert@calpoly.edu}
\author[0000-0001-9428-6238]{Nico~Winkel}
\affiliation{Max-Planck-Institut f\"ur Astronomie, K\"onigstuhl 17,
  D-69117 Heidelberg, Germany}
\email{winkel@mpia.de}
\author[0000-0002-8460-0390]{Tommaso~Treu}
\affiliation{Department of Physics and Astronomy, University of California, Los Angeles, 430 Portola Plaza, Los Angeles, CA 90095, USA}
\email{tt@astro.ucla.edu}
\author[0000-0001-8917-2148]{Xuheng~Ding}
\affiliation{School of Physics and Technology, Wuhan University, Wuhan 430072, China}
\email{dingxuheng@126.com}
\author[0000-0002-1912-0024]{Vivian~U}
\affiliation{IPAC, California Institute of Technology, 1200 E. California Blvd., Pasadena, CA 91125, USA}
\affiliation{Department of Physics and Astronomy, 4129 Frederick Reines Hall, University of California, Irvine, CA 92697, USA}
\email{vivianu@uci.edu}
\author[0000-0002-0164-8795]{Raymond~P.~Remigio}
\affiliation{Department of Physics and Astronomy, 4129 Frederick Reines Hall, University of California, Irvine, CA 92697, USA}
\email{remigior@uci.edu}
\author[0000-0002-3026-0562]{Aaron~J.~Barth}
\affiliation{Department of Physics and Astronomy, 4129 Frederick Reines Hall, University of California, Irvine, CA 92697, USA}
\email{barth@uci.edu}
\author[0000-0001-6919-1237]{Matthew~A.~Malkan}
\affiliation{Department of Physics and Astronomy, University of California, Los Angeles, 430 Portola Plaza, Los Angeles, CA 90095, USA}
\email{malkan@astro.ucla.edu}
\author[0000-0002-1961-6361]{Lizvette~Villafa{\~n}a}
\affiliation{Physics Department, California Polytechnic State University, San Luis Obispo, CA 93407, USA}
\email{lvillafa@calpoly.edu}
\author{Samantha Allen}
\affiliation{Physics Department, California Polytechnic State University, San Luis Obispo, CA 93407, USA}
\email{samanthakallen@rocketmail.com}
\author{Ellie Johnson}
\affiliation{Physics Department, California Polytechnic State University, San Luis Obispo, CA 93407, USA}
\email{ejohn127@calpoly.edu}
\author{Sebastian Contreras}
\affiliation{Physics Department, California Polytechnic State University, San Luis Obispo, CA 93407, USA}
\affiliation{Department of Astronomy, San Diego State University, San Diego, CA 92182, USA}

\email{scontreras0466@sdsu.edu}
\author[0000-0002-3560-0781]{Minjin Kim}
\affiliation{Kyungpook National University, Daegu 41566, Republic of Korea}
\affiliation{Department of Astronomy, Yonsei University, 50 Yonsei-ro, Seodaemun-gu, Seoul 03722, Republic of Korea}
\email{mkim.astro@gmail.com}
\author[0000-0003-3195-5507]{Simon~Birrer}
\affiliation{Department of Physics and Astronomy, Stony Brook University, Stony Brook, NY 11794, USA}
\email{simon.birrer@stonybrook.edu}
\author[0000-0003-3804-2137]{Knud~Jahnke}
\affiliation{Max-Planck-Institut f\"ur Astronomie, K\"onigstuhl 17, D-69117 Heidelberg, Germany}
\email{jahnke@mpia.de}
\author[0000-0001-8917-2148]{Shaoping~Zheng}
\affiliation{School of Physics and Technology, Wuhan University, Wuhan 430072, China}
\email{spzheng@whu.edu.cn}
\correspondingauthor{Vardha N.~Bennert}
\email{vbennert@calpoly.edu}

\begin{abstract}
Reverberation mapping (RM) determines the mass of black holes (BH) in active galactic nuclei (AGNs) by
resolving the BH gravitational sphere of influence in the time domain. Recent 
RM campaigns yielded direct BH masses through dynamical modeling for a sample of 32 objects, spanning a wide
range of AGN luminosities and BH masses. In addition, 
accurate BH masses have been determined by spatially resolving the broad-line region with GRAVITY for a handful of AGNs. 
Here, we present a detailed analysis of {\it Hubble Space Telescope}
images using surface-brightness profile fitting with state-of-the-art programs.
We derive AGN luminosity and host-galaxy properties, such as radii and
luminosities for spheroid, disk, and bar (if present).
The spheroid effective radii
were used to measure stellar velocity dispersion from integral-field spectroscopy.
Since the BH masses of our sample do not depend on any assumption of the
virial factor needed in single-epoch spectroscopic mass estimates, we can show that the resulting
scaling relations between the mass of the supermassive BHs and their host
galaxies match those of quiescent galaxies, naturally extending to lower masses
in these (predominantly) spiral  galaxies. We find that the inner AGN orientation, as traced by the broad-line region inclination angle, is uncorrelated 
with the host-galaxy disk.
Our sample has the most direct and accurate \mbh~measurements
of any AGN sample and provides a fundamental local benchmark for studies of the
evolution of massive black holes and their host galaxies across cosmic time. 
\end{abstract}


\section{Introduction} \label{sec:intro}
Supermassive black holes (SMBHs) are common in the centers of
massive galaxies. Their masses (\mbh) correlate
with the properties of their host galaxies,
such as stellar mass, spheroid luminosity, and
stellar velocity dispersion \sigstar~\citep[e.g.,][]{Magorrian:1998, Gebhardt:2000, Ferrarese_Merritt:2000}.
The tightness of the \mbh--host galaxy relations ($\sim$0.4 dex) over a large
range of masses indicates that
 the formation and evolution of galaxies is linked to that of black holes (BHs).
Mechanisms explaining the co-evolution of SMBHs and their host galaxies involve feedback from Active Galactic Nuclei 
\citep[AGNs; e.g.,][]{DiMatteo:2005,Fabian:2012,Hopkins:2016,Morganti:2017} and hierarchical growth
through major mergers \citep[][]{Peng:2007,Jahnke:2011}.
Despite many  observational studies,
both in the local universe
and across cosmic history
\citep[for reviews, see, e.g.,][]{Ferrarese_Ford:2005, Kormendy_Ho:2013, Heckman_Best:2014, Graham:2016},
open questions remain, such as the origin and drivers of the
relations, the role of host-galaxy
morphology, and the evolution of the relations with cosmic time. 
To make progress, it is essential to understand systematic
uncertainties, and measure both \mbh~and host properties as
robustly as possible in the local universe, which serves as
an anchor for high-redshift studies.

In local quiescent galaxies, the gravitational sphere of influence of the SMBH can be spatially resolved 
by tracing the motion of stars and gas in the center, resulting in an
estimate of \mbh.
For more distant galaxies, reverberation mapping (RM) of  accreting
SMBHs in type-1 AGNs resolves the gravitational sphere of influence of the BH in the time domain
\citep[e.g.,][]{Wandel:1999,Vestergaard:2002}, 
using gas in the vicinity of the BH, 
the broad-line region (BLR).
The BLR gas is photoionized by the AGN accretion disk continuum. It emits
Doppler-broadened lines (e.g., Hydrogen) upon recombination,  the velocity of their broadening
assumed to be accelerated primarily by the gravity of  
the nearby SMBH.
In RM,  the time-delayed response of the BLR to ionizing radiation from the
central accretion disk
is monitored. The time lag between ionizing continuum changes and corresponding BLR changes measures the light-crossing time and thus, the characteristic size of the BLR. When combined with
the velocity of the gas, as traced by the broad emission line
width, \mbh~can be estimated up to the ``virial'' factor $f$ that
depends on the unknown geometry and kinematic structure of the BLR:
$M_{\rm BH} = f R_{\rm BLR} V^2 / G$ 
($f$ = virial factor, 
$R_{\rm BLR}$ = BLR radius, 
$V$ = broad-line width, and $G$ = gravitational constant).

A sample-averaged $f$ value has typically been estimated by matching the \mbh--\sigstar~relation of active galaxies 
to that of quiescent galaxies
\citep[e.g.,][]{Onken:2004,Woo:2010,Park:2012}.
The importance of the AGN sample with \mbh~based on RM (the ``RM sample'')
lies in determining an empirical relation between BLR size and accretion-disk luminosity \citep[e.g.,][]{Bentz:2013}.
This size-luminosity relation enables rough estimates of single-epoch \mbh~masses of 
hundreds of thousands of AGNs across cosmic history
\citep[e.g.,][]{Shen:2011,Willott:2010,Mortlock:2011,Rakshit:2020},
to study the cosmic evolution of the \mbh~scaling
relations \citep[e.g.,][]{Treu:2004,Bennert:2010,Ding:2021a},
and has been used in recent JWST discoveries of SMBHs
at very high redshift
\citep[e.g.,][]{Pacucci:2024,Greene:2024}.

Recent progress has enabled direct \mbh~measurements 
by forward modeling the BLR emission.
Our team has been using the 
Code for AGN Reverberation and Modeling of Emission Lines (\texttt{CARAMEL})
\citep[e.g.,][]{Brewer:2011,Pancoast:2011,Pancoast:2014,Pancoast:2018,Williams:2018,Villafana:2022,Bentz:2022,Bentz:2023a,Bentz:2023b}.
Based on a Bayesian framework, \texttt{CARAMEL} explores a
27-parameter space that incorporates \mbh~and various geometric and dynamical properties of the BLR. It determines the combination of model parameters
that best fits the RM data, creating a 3D map of the high-speed gas
that surrounds the BH.
Importantly, this method provides the most precise \mbh~in AGNs,
without the need of invoking the virial factor.
Note that other teams have developed similar codes for the same purpose, such as BRAINS \citep{Li:2013,Li:2018,Stone:2025} and GRAVITY
\citep{GravityCollaboration:2018,GravityCollaboration:2021a,GravityCollaboration:2021b}.

The CARAMEL sample has grown to 32 AGNs,
spanning a wide range of luminosities and BH masses (6.4$<$log \mbh/\msun$<$9.1).
In addition, for a handful of AGNs, 
accurate \mbh~has been determined by spatially resolving (and modeling) the BLR with 
GRAVITY at the Very Large Telescope Interferometer
\citep[VLTI][]{GravityCollaboration:2018,GravityCollaboration:2021a,GravityCollaboration:2021b}.

Of the observed \mbh--host galaxy scaling relations, the one between BH
mass and stellar velocity dispersion, \mbh--\sigstar, is of particular interest,
since it has been used to determine the sample-averaged
virial factor for RM AGNs.
However, it is known that \sigstar~can vary significantly
across definitions common in the literature.
Measurements 
depend on the aperture used,
the effects of inclination, the presence of a kinematically cold, rotating
disk and, for type-1 AGN, the non-stellar continuum outshining the
underlying stellar continuum and its absorption lines
\citep[e.g.,][]{Debattista:2013, Hartmann:2014, Bennert:2015}.
These uncertainties caution
the use of \sigstar~determined from long-slit spectroscopy or
fibers.
Integral-field
unit (IFU)  spectroscopy is the best way forward towards a precise
measurement of \sigstar, using spatially resolved kinematics and
integrating over the effective radius.

Our team obtained IFU data
with Keck/KCWI and VLT/MUSE
for the RM sample with directly measured \mbh~\citep[][]{Winkel:2025a,Remigio:2025}.
The IFU observations map stellar
kinematics across the galaxy and spheroid effective radii.
After accounting for different \mbh~distributions, we can demonstrate
directly that AGNs
follow the same \mbh--\sigstar~relation as quiescent galaxies.
Fitting the derived \mbh--\sigstar~relation
yields the same virial factor $f$ as from the classical method, as well as the
average of individual $f$-factors from dynamical modeling
\citep[log $f_{\rm dyn}$ = 0.66$\pm$0.07,][]{Winkel:2025a}.
The \mbh--\sigstar~relation is robust, regardless of host-galaxy morphology, and
there is no significant dependence of the sample-average $f$-factor on
host-galaxy morphology \citep{Winkel:2025a}.

The stellar velocity dispersion measurements presented in
\citet{Winkel:2025a} are integrated over effective radii of the spheroid and/or total host galaxy. 
In this paper, we describe
the detailed analysis of the AGN host galaxies that led
to the measurements of these effective radii (and thus stellar-velocity
dispersion).
We present high
resolution and high signal-to-noise (S/N) Hubble Space Telescope (HST)
images of the RM and GRAVITY sample with directly
measured \mbh.
 The image analysis uses state-of-the-art
programs (\texttt{galight} and \texttt{lenstronomy}) to
derive host-galaxy properties, such as luminosities,
radii, and morphologies.
When combined with the directly measured BH masses,
we can create the full range of \mbh--host galaxy scaling relations.
Using BLR modeling results, we can compare the inclination
between BLR and host-galaxy disk, to study their relative orientations which provides clues to SMBH feeding.

The paper is organized as follows.
Sample selection is summarized in Section~\ref{sec:sample}.
HST observations and data reduction
are presented in Section~\ref{sec:obs}.
Surface-brightness profile modeling is described in
Section~\ref{sec:sbp}.
Results are discussed in Section~\ref{sec:results}.
The paper concludes with a summary and outlook (Section~\ref{sec:summary}).
Appendix \ref{Appendix:Lenstronomy} presents details of 
surface-brightness fitting results. 

Throughout this paper, we adopt $H_0$ = 67.8 \kms~\perMpc,
$\Omega_m$ = 0.308, and $\Omega_{\Lambda}$ = 0.692
\citep{Planck:2016}.

\section{Sample Selection and BH Masses}
\label{sec:sample}
The sample consists of 32 local RM AGNs with directly measured \mbh~from
dynamical modeling \citep[Table~\ref{tab:results}; for details, 
see Table~2 in][]{Villafana:2023}.
\footnote{Note that for three of these AGNs, Ark\,120, Mrk\,110, and Mrk\,142,
BLR models fit the data only with moderate quality \citep{Villafana:2022}, and have not  been further included in CARAMEL-extended sample papers. We include these objects here, but our results do not change if we exclude them.}
This sample has the most robust \mbh~measurements in AGNs,
as it does not rely on the assumption of a virial factor.
The low-redshift
(0.01 $\le z \le 0.16$) 
sample covers a large range in AGN luminosities and BH masses
\citep[log $L_{\rm bol}/{\rm erg\,s}^{-1} \sim 10^{43-47}$, 
log \mbh/M$_{\odot}$ $\sim$ 6.4--8.3,][]{Winkel:2025a}.
We include five AGNs 
for which accurate \mbh~have been determined by spatially resolving the BLR with GRAVITY. Two of these five GRAVITY AGNs have both CARAMEL RM and GRAVITY measurements: NGC 3783 and IC 4329A.
Note that while the GRAVITY collaboration has published results for a
total of seven local AGNs so far, we include only those with existing
IFU and HST data.

Finally, we include nine AGNs for which the BH mass has been
determined through classical RM (cRM), but for which the data were not of sufficient
quality for dynamical modeling. For these objects,  we adopt a
virial factor of log $f$ = 0.65 to re-calculate \mbh. This value is consistent with the average of the individual values of log $f_{\rm dyn}$ = 0.66$\pm$0.07 determined in \citet{Winkel:2025a}
\citep[see also][]{Villafana:2023}. 
The final sample thus consists of 44 objects.
The most important sample parameters are listed 
in Tables~\ref{tab:hst} and
\ref{tab:results};
further details are given by \citet{Winkel:2025a}.
As our comparison sample, we use 
51 quiescent galaxies (either ellipticals or spirals/S0 with classical bulges) with \mbh\ based on dynamical modeling of spatially
resolved stellar kinematics from \citet[hereafter KH13]{Kormendy_Ho:2013}.
For details on the quiescent sample, see Section 3.4 in \citet{Bennert:2021}.
KH13 lists
spheroid magnitudes in the $V$ band as well as colors, which were used to calculate the $I$-band luminosities
as described by \citet{Bennert:2021}.

\startlongtable
\begin{deluxetable*}{lccccccc}
\tabletypesize{\footnotesize}
\tablecolumns{8}
\tablecaption{Sample Properties and HST Observations}
\tablehead{\colhead{AGN Name} & 
\colhead{R.A.} &
\colhead{Decl.} &
\colhead{$z$} &
\colhead{Instrument} &
\colhead{Filter} &
\colhead{Program ID} &
\colhead{Reference}
                            \\
& (J2000) & (J2000) & & & & \\
(1) & (2)  & (3) & (4) & (5) & (6) & (7)  & (8)  }
\startdata
Mrk 335 & 00:06:19.52 & +20:12:10.5 & 0.0258 & WFC3/UVIS & F814W & 17103 & B26 \\
Mrk 1501 & 00:10:31.01 & +10:58:29.5 & 0.087 & ACS/WFC1 & F814W & 15444 & K21 \\
Zw 535-012 & 00:36:20.93 & +45:39:53.7 & 0.0476 & ACS/WFC1 & F814W & 15444 & K21 \\
Mrk 590 & 02:14:33.56 & $-$00:46:00.2 & 0.0261 & ACS/HRC & F550M & 9851 & B06 \\
Mrk 1044 & 02:30:05.52 & $-$08:59:53.2 & 0.0161 & ACS/WFC1 & F814W & 15444 & K21 \\
Mrk 1048 & 02:34:37.77 & $-$08:47:15.4 & 0.0427 & ACS/WFC1 & F814W & 15444 & K21 \\
3C 120 & 04:33:11.10 & +05:21:15.6 & 0.033 & ACS/HRC & F550M & 9851 & B06 \\
Ark 120 & 05:16:11.42 & $-$00:08:59.4 & 0.0327 & ACS/WFC1 & F814W & 15444 & K21 \\
NGC 2617 & 08:35:38.80 & $-$04:05:17.9 & 0.0142 & WFC3/UVIS & F547M & 13816 & B18 \\
IRAS 09149-6206 & 09:16:09.39 & $-$62:19:29.9 & 0.0573 & WFC3/UVIS & F814W & 17103 & B26 \\
MCG +04-22-042 & 09:23:43.00 & +22:54:32.6 & 0.0332 & ACS/WFC1 & F814W & 15444 & K21 \\
Mrk 110 & 09:25:12.85 & +52:17:10.4 & 0.0355 & ACS/HRC & F550M & 9851 & B06 \\
Mrk 1239 & 09:52:19.16 & $-$01:36:44.1 & 0.0196 & WFPC2/PC1 & F606W & 5479 & M98 \\
Mrk 141 & 10:19:12.56 & +63:58:02.8 & 0.0417 & WFC3/UVIS & F814W & 17103 & B26 \\
NGC 3227 & 10:23:30.58 & +19:51:54.3 & 0.0038 & WFC3/UVIS & F547M & 11661 & B18 \\
Mrk 142 & 10:25:31.28 & +51:40:34.9 & 0.0446 & WFC3/UVIS & F547M & 11662 & B13 \\
NGC 3516 & 11:06:47.49 & +72:34:06.9 & 0.0088 & WFC3/UVIS & F547M & 11661 & B18 \\
SBS 1116+583A & 11:18:57.69 & +58:03:23.7 & 0.0279 & WFC3/UVIS & F547M & 11662 & B13 \\
Arp 151 & 11:25:36.17 & +54:22:57.0 & 0.0207 & WFC3/UVIS & F547M & 11662 & B13 \\
NGC 3783 & 11:39:01.76 & $-$37:44:19.2 & 0.0097 & WFC3/UVIS & F547M & 11661 & B18 \\
Mrk 1310 & 12:01:14.36 & $-$03:40:41.1 & 0.0195 & WFC3/UVIS & F547M & 11662 & B13 \\
NGC 4151 & 12:10:32.58 & +39:24:20.6 & 0.0033 & WFC3/UVIS & F547M & 11661 & B18 \\
PG 1211+143 & 12:14:17.67 & +14:03:13.2 & 0.081 & ACS/HRC & F550M & 10833 & B09a \\
Mrk 50 & 12:23:24.14 & +02:40:44.8 & 0.0234 & WFC3/UVIS & F814W & 16014 & B26 \\
NGC 4593 & 12:39:39.43 & $-$05:20:39.3 & 0.0083 & WFC3/UVIS & F547M & 11661 & B18 \\
PG 1310-108 & 13:13:05.79 & $-$11:07:42.4 & 0.0343 & WFC3/UVIS & F814W & 17103 & B26 \\
RBS 1303 & 13:41:12.90 & $-$14:38:40.6 & 0.0418 & WFC3/UVIS & F814W & 17103 & B26 \\
IC 4329A & 13:49:19.26 & $-$30:18:34.2 & 0.0151 & ACS/HRC & F550M & 10516 & B09a \\
Mrk 279 & 13:53:03.45 & +69:18:29.6 & 0.0305 & WFC3/UVIS & F814W & 17103 & B26 \\
NGC 5548 & 14:17:59.54 & +25:08:12.6 & 0.0163 & WFC3/UVIS & F547M & 11661 & B18 \\
PG 1426+015 & 14:29:06.57 & +01:17:06.2 & 0.086 & WFPC2/PC1 & F547M & 10833 & B09a \\
Mrk 841 & 15:04:01.20 & +10:26:16.2 & 0.0364 & WFC3/UVIS & F814W & 17103 & B26 \\
Mrk 1392 & 15:05:56.55 & +03:42:26.3 & 0.0359 & WFC3/UVIS & F814W & 17103 & B26 \\
Mrk 1511 & 15:31:18.07 & +07:27:27.9 & 0.0339 & WFC3/UVIS & F547M & 13816 & B18 \\
PG 1617+175 & 16:20:11.28 & +17:24:27.5 & 0.112 & WFPC2/PC1 & F547M & 10833 & B09a \\
NPM 1G+27.0587 & 18:53:03.87 & +27:50:27.7 & 0.062 & WFC3/UVIS & F814W & 17103 & B26 \\
Zw 229-015 & 19:05:25.94 & +42:27:39.7 & 0.0279 & ACS/WFC1 & F814W & 15444 & K21 \\
NGC 6814 & 19:42:40.64 & $-$10:19:24.6 & 0.0052 & WFC3/UVIS & F814W & 12961 & B19 \\
RXJ 2044.0+2833 & 20:44:04.50 & +28:33:12.1 & 0.05 & ACS/WFC1 & F814W & 15444 & K21 \\
Mrk 509 & 20:44:09.75 & $-$10:43:24.7 & 0.0347 & WFC3/UVIS & F547M & 12212 & F13 \\
PG 2130+099 & 21:32:27.81 & +10:08:19.5 & 0.064 & ACS/HRC & F550M & 9851 & B06 \\
PG 2209+184 & 22:11:53.89 & +18:41:49.9 & 0.07 & WFC3/UVIS & F814W & 17103 & B26 \\
RBS 1917 & 22:56:36.50 & +05:25:17.2 & 0.065 & WFC3/UVIS & F814W & 17103 & B26 \\
NGC 7469 & 23:03:15.67  & +08:52:25.3 & 0.0166 & WFC3/UVIS & F547M & 11661 & B18 \\
\enddata
\tablecomments{
Col. (1): AGN Name.		      	   	  
Col. (2): Right ascension in hours, minutes and seconds (NED).				  
Col. (3): Declination in degrees, arcminutes and arcseconds (NED).
Col. (4): Redshift (NED).
Col. (5): Instrument and aperture.
Col. (6): Filter.
Col. (7): Program ID.
Col. (8): Original paper: 
B08: \citet{Bennert:2008};
B26: this paper;
B06: \citet{Bentz:2006};
B09a: \citet{Bentz:2009a};
B13: \citet{Bentz:2013};
B18: \citet{Bentz:2018};
C07: \citet{Canalizo:2007};
F13: \citet{Fischer:2013}
K21: \citet{Kim:2021};
M98: \citet{Malkan:1998}.
}
            \label{tab:hst}
\end{deluxetable*}

\section{Hubble Space Telescope Observations and Data Reduction}
\label{sec:obs}
The HST data are a combination of
newly conducted observations (PI Bennert; GO 17103;
11 objects), as well as archival images
obtained from different programs.
For program GO 17103, each object was observed for one orbit with WFC3
in filter F814W.
Long exposures (500--600\,s) were complemented with
short, unsaturated ones (10\,s), for a full dynamic range.
The observing sequence (short, long, long -- dither -- long, long,
short) minimizes buffer dump, using a manual dither pattern corresponding to the 
``WFC3-UVIS-GAPLINE''.
Archival images were obtained using WFPC2, ACS or WFC3 in medium
or broad band $V$ or $I$ filters.
Details on the observations are given in Table~\ref{tab:hst}.

All data are reduced and analyzed in
a homogeneous way,
using the standard HST calibration pipeline.
In addition, \texttt{LA-Cosmic} \citep{vanDokkum:2001} was run before
image combination to improve cosmic ray rejection.
For those objects for which the central AGN point source was saturated in the long
exposures, short exposures,
scaled by exposure time, were used to replace the saturated pixels.
Long exposures were dither-combined using the package \texttt{AstroDrizzle}.
Various drizzle parameters (e.g., scale and pixfrac) were applied to
find the best settings,
based on the resulting resolution and image quality.
In the ACS archival dataset (GO 15444), the short-exposure sub-array images were affected by bias striping noise, which was mitigated using the \texttt{acs\_destripe\_plus} task
\citep{Grogin:2010,Grogin:2011,Kim:2021}.
Resulting images are of high quality, both in terms of resolution and S/N.

\section{Host Galaxy Surface-Brightness Fitting}
\label{sec:sbp}
HST images were analyzed 
using the 2D host-galaxy fitting software \texttt{lenstronomy}
\citep{Birrer_Amara:2018,Birrer:2021}, in combination with \texttt{galight} \citep{Ding:2020}.
\texttt{Lenstronomy} is a modern version of \texttt{GALFIT}
\citep{Peng:2002}, allowing for a more
general surface-brightness reconstruction
with coefficients determined through 
linear minimization rather than a non-linear parameter fitting. 
The Markov Chain Monte Carlo (MCMC) technique provides realistic errors and explores the
covariance between various model parameters.
\texttt{Lenstronomy} was designed for galaxy lensing, but can be applied
broadly for general 2D galaxy decompositions. The
source code is public (see Astrophysics Source Code Library) and can
be directly modified by the user.
The \texttt{galight} software package offers a user-friendly application of \texttt{lenstronomy} 
and includes helpful tools to extract Point Spread Function (PSF) stars, create masks, and
subtract the background. 

For each object, the full data analysis includes the following steps
\citep[see also][for further details on the general procedure]{Bennert:2021}:

(1) PSF stars are chosen from stars in the field of view
(FOV). Ideally, these stars are bright, but not saturated, and close
to the AGN location in the FOV. A library of PSF stars is created from
suitable stars in all available exposures. Multiple PSF stars from
different images (obtained with the same camera, in the same filter, at the same orientation and close in time) are combined into one high S/N PSF, 
through an iterative PSF reconstruction \citep[e.g.,][available on github as \texttt{psfr}]{Birrer:2019,Shajib:2019,Shajib:2020}.

(2) The final-reduced image of the AGN host galaxy is 
examined to determine morphology and components fitted (spheroid, disk,
bar). For objects for which there is at least an indication of a
visual bar, the bar parameters size, position angle, and ellipticity are
determined manually and used as starting parameters for the
bar in the spheroid--disk--bar fit.

(3) The sky background is subtracted.

(4) A noise-level map is created, taking into account the dithering
process.

(5) A spherical image cut-out is created, centered on the AGN.
Nearby objects are masked or fitted simultaneously, depending on
brightness and proximity to, or overlap with, the AGN host galaxy.

(6) For each object, the central AGN is fitted with a point source (PSF).
The host galaxy is fitted with three different models: (a) a
spheroid-only component (free S{\'e}rsic index with 1 $\le$ $n$ $\le$ 5); (b) a spheroid plus disk
component (S{\'e}rsic index $n$ = 1; if present); (c) a spheroid plus disk
plus bar component (elliptical S{\'e}rsic model with $n$ = 0.5; if present).
Based on pre-defined starting parameters and constraints,
\texttt{lenstronomy} 
uses a Particle Swarm Optimizer (PSO) as a minimizer
and an MCMC sampler to
calculate the posterior distribution of the parameter space. 
We experiment with different
settings for the PSO (number of particles and iterations). For all
fits, we plot the PSO particle positions, parameter velocity and
inferred flux as a function of iteration to ensure that the chain
converged.

(7) We first run the spheroid-only fit. This fit is robust and gives
the same results independent of starting parameters used.

(8) We then use the results from the spheroid-only fit as starting
parameters for the disk in the subsequent spheroid--disk and
spheroid--disk--bar fit (if required).
(a)
For the spheroid--disk fit, we
force the disk to be larger than the spheroid and more
elliptical.
Three different starting parameters are used for
both the spheroid and disk effective radius (see
Table~\ref{tab:starting}).
In combination, this results in nine different models that are run, ensuring a true global
minimum is reached in the best fit.
(b) For the spheroid--disk--bar fit, we force the disk
to be larger than the bar and the spheroid, and the spheroid component
to be the most round one of three components.
The same starting parameters are used
for spheroid and disk as above (8a), with the bar added with specific starting
parameters (radius and position angle) determined visually.

(9) Through careful inspection and comparison of the fits for the
different runs (such as residual plots, final results, chi–square
values, uncertainties), we  determine the best model and fit for all
objects.

(10) The fits are used to distinguish between 
classical and pseudo-bulges.
We use the S{\'e}rsic index as a first indicator, but
conservatively, take into account other parameters as well (presence
of bar, inclination, spheroid-to-total luminosity ratio, rotation;
discussed in detail in Section\ref{sec:morph}).

(11) Galactic foreground extinction is subtracted using the 
dust extinction map \citep{Schlafly_Finkbeiner:2011}.

(12) Standard cosmology is  applied to convert arcsecond radii to
parsec and apparent magnitudes to luminosities. The
$k$-correction uses \texttt{PySynphot}
and an Sa galaxy template \citep{Kinney:1996}.
The F814W filter magnitudes are virtually identical to $I$-band
magnitudes
\citep{Harris:2018}, given the colors of all galaxy templates.
Based on the different fitting results (using different starting parameters and PSFs etc.), 
we conservatively adopt 0.04 dex as uncertainty on the
derived luminosities.

 General notes on the modelling procedure and results: 
(1) The central AGN point source was fitted with a PSF. (a) While the HST PSF has a full width at half maximum (FWHM) of $\sim$2.5 pixels, it extends outwards by several arcseconds and includes diffraction spikes and sometime ghosts. However, 98\% of the encircled energy is within 2 arcseconds
\citep[for details, see Sections 6.6.1--6.6.3 in the WFC3 Instrument Handbook;][]{Pagul:2024}.
(b) In some sources, the AGN position is offset from the centroid of the host-galaxy spheroid component by 0\farcs1--0\farcs2. Such offsets can be caused by central dust lanes, which are commonly seen in these type-1 Seyfert galaxies, or by the intrinsic asymmetry common in AGN hosts, and due to recent mergers or interactions \citep[for discussions, see, e.g.,][]{Kim:2017}.
(2) In many objects, residuals of rings and spiral arms remain after the fitting. This is a natural consequence of fitting 
intrinsically asymmetric features such as spiral arms, or asymmetries caused by mergers or interactions with
spherically symmetric models with a limited number of components (i.e., AGN, spheroid, disk, bar, if present). While increasing the number of components and/or including special functions 
(such as a hyperbolic tangent rotation function) with Fourier modes would allow us to fit these asymmetries 
and measure asymmetry parameters
\citep{Kim:2017}, 
this is beyond the scope of our paper.
Physical quantities relevant to this paper, such as the radii of the spheroid and disk as well as the magnitudes can be robustly inferred by the physically-motivated models we chose.
 (3) For NGC\,4593, the HST FOV is too small and mostly covers the bar. While spheroid and bar component are reliably fitted, the disk fit (which is much larger than the FOV) is unreliable.
 (4) For Mrk\,279, the bottom chip was affected by a strong ``ghost'' and thus masked out and excluded from the fitting.

\begin{deluxetable*}{cc}
\tabletypesize{\footnotesize}
\tablecolumns{2}
\tablecaption{Starting Parameters for Spheroid-Disk Fit}
\tablehead{
\colhead{Disk effective radius $R_{\rm disk}$} &
\colhead{Spheroid effective radius $R_{\rm sph}$} \\
(1) & (2) }
\startdata
$R_{\rm SS}$ & 0.2 $R_{\rm SS}$\\
 & 0.5 $R_{\rm SS}$\\
 & 0.75 $R_{\rm SS}$\\
\hline
 0.5 $R_{\rm SS}$ & 0.1 $R_{\rm SS}$\\
 & 0.25 $R_{\rm SS}$\\
 & 0.375 $R_{\rm SS}$\\
\hline
 2 $R_{\rm SS}$ & 0.4 $R_{\rm SS}$\\
 & $R_{\rm SS}$\\
 & 1.5 $R_{\rm SS}$\\
\hline
\enddata
\tablecomments{
\label{tab:starting}
To ensure a true global minimum is reached, 
we choose a series of nine different starting parameters for the effective radii of the spheroid+disk fit, 
relative to the effective radius 
derived from the best fitting
single S{\'e}rsic profile $R_{\rm SS}$
from the spheroid-only fit.
For the spheroid+disk+bar fit, the same combination is used, with the bar added with specific starting parameters (radius and position angle) as determined visually.
Col. (1): Starting parameters for disk effective radius $R_{\rm disk}$.      	   	  
Col. (2): Starting parameters for spheroid effective radius $R_{\rm sph}$.
}
\end{deluxetable*}

\section{Results and Discussion}
\label{sec:results}

\subsection{Host Galaxy Morphology}
\label{sec:morph}
By visual inspection of the images and residuals of the 
surface-brightness fitting,
host-galaxy morphologies were determined.
Results of the surface-brightness fitting are shown in 
Appendix~\ref{Appendix:Lenstronomy}.
Table~\ref{tab:results}, column 7, lists the host-galaxy morphology.

Of our sample of 44 AGNs,
only five objects (11\%) are hosted by bona-fide elliptical galaxies.
The majority of AGNs (39/44; 89\%) are hosted by spiral or S0 galaxies.
Of the disk galaxies, 18/39 (46\%) were best fitted with an additional
bar component. 
Seyfert galaxies are generally considered to have similar bar
fractions as quiescent spiral galaxies \citep[see, e.g.,][]{Hunt:1999}: 1/3 no bar, 1/3 weak bar
(``SAB''), and 1/3 with a strong bar (``SB''). Thus,
we are likely missing a small fraction of weak bars due to dust obscuration in the optical, especially in highly inclined disks.

Two objects (5\%) show strong
signs of merger activity, Mrk\,1048 and Arp\,151.
The HST image of Mrk\,1048 reveals two nuclei and a tidal tail forming a central ring structure; the latter had been noted previously and interpreted as evidence of merger activity \citep{Parker:2014}.
The HST image of Arp\,151 shows a highly elongated one-sided tidal tail-like structure with a second nucleus.
A handful of other objects
(such as Mrk\,1501, 3C\,120, Ark\,120, IRAS\, 09149-6206, PG\,1310-108, Mrk\,279, NPM\,1G+27.0587)
show asymmetries,
nearby neighbors, or tidal tails.
Overall, the distribution of host-galaxy morphologies is typical for
Seyfert-type AGNs \citep[e.g.,][]{Kocevski:2012,
 Schawinski:2012, Marian:2019, Husemann:2022,Bennert:2021,Kim:2017,Kim:2021},
and also low-redshift PG quasars \citep[e.g.,][]{Zhao:2021}.

For all galaxies fitted by either spheroid+disk or spheroid+disk+bar,
we classify the spheroid as pseudo-bulge, if at least three
of the following four criteria are met
\citep{Kormendy_Ho:2013,Bennert:2021}:
(1) S{\'e}rsic index $<$ 2;
(2) spheroid-to-total luminosity ratio $<$ 0.5;
(3) rotation dominated, i.e., ratio between maximum rotational
velocity (corrected for inclination) at effective spheroid radius and central stellar-velocity
dispersion $>$ 1;
(4) for face-on galaxies, the presence of a bar is considered an
indicator for the existence of a pseudo-bulge.
Of the late-type galaxies, 23/40 (58\%) have  been found to harbor
pseudo-bulges in this way (see Table~\ref{tab:results}).
This fraction is similar to a sample of both local type-1 AGNs and low-redshift PG quasars \citep{Bennert:2021, Zhao:2021}.

\startlongtable
\begin{deluxetable*}{lccccccccccc}
\tabletypesize{\footnotesize}
\tablecolumns{12}
\tablecaption{Black Hole Masses and Derived Host Galaxy Properties}
\tablehead{\colhead{AGN Name} & 
  \colhead{\mbh} &
 \colhead{$L_{\rm{sph}}$} &
  \colhead{$L_{\rm{sph+bar}}$} &
     \colhead{$L_{\rm{disk}}$} &
\colhead{$L_{\rm{host}}$} &
\colhead{Host} &
\colhead{$i_{\rm host}$} &
\colhead{$i_{\rm BLR}$} &
\colhead{$o_{\rm BLR}$} &
\colhead{Sample} &
\colhead{Reference}
\\
& (log \msun) &
  (log \lsun) &
  (log \lsun) &
  (log \lsun) &
  (log \lsun) &
  & ($\degr$)
    & ($\degr$) & ($\degr$) \\
(1) & (2)  & (3) & (4) & (5) & (6) & (7) & (8) & (9) & (10) & (11) & (12)}
\startdata
Mrk 335 & 7.25$^{+0.1}_{-0.1}$ & 9.86 & \nodata & 9.65 & 10.07 & BD(C) & 25 & 35$^{+4}_{-5}$ & 38$^{+5}_{-5}$ & CARAMEL & G17 \\
Mrk 1501 & 7.86$^{+0.2}_{-0.17}$ & 10.56 & \nodata & 10.49 & 10.82 & BD(C) & 44 & 20$^{+5}_{-6}$ & 22$^{+12}_{-6}$ & CARAMEL & G17 \\
Zw 535-012 & 7.57$^{+0.15}_{-0.1}$ & 9.79 & 10.39 & 10.20 & 10.62 & BDB(P) & 60 & \nodata & \nodata & cRM & U22$^{\ast\ast}$ \\
Mrk 590 & 7.58$^{+0.07}_{-0.07}$ & 9.83 & \nodata & 9.85 & 10.14 & BD(C) & 42 & \nodata & \nodata & cRM & M21$^{\ast\ast}$ \\
Mrk 1044 & 6.1$^{+0.12}_{-0.1}$ & 9.46 & 9.87 & 9.66 & 10.09 & BDB(P) & 29 & \nodata & \nodata & cRM & D15$^{\ast\ast}$ \\
Mrk 1048 & 7.79$^{+0.44}_{-0.48}$ & 10.35 & \nodata & 10.83 & 10.95 & BD(P) & 48 & 22$^{+9}_{-9}$ & 31$^{+14}_{-10}$ & CARAMEL & V22 \\
3C 120 & 7.84$^{+0.14}_{-0.19}$ & 9.64 & \nodata & \nodata & 9.64 & B(C) & \nodata & 18$^{+5}_{-3}$ & 21$^{+8}_{-5}$ & CARAMEL & G17 \\
Ark 120 & 8.26$^{+0.12}_{-0.17}$ & 10.63 & \nodata & 10.24 & 10.78 & BD(C) & 31 & 14$^{+4}_{-3}$ & 32$^{+7}_{-8}$ & CARAMEL$^1$ & V22 \\
NGC 2617 & 7.51$^{+0.47}_{-0.47}$ & 9.12 & \nodata & 10.09 & 10.13 & BD(C) & 19 & \nodata & \nodata & cRM & F17$^{\ast\ast}$ \\
IRAS 09149-6206 & 8.0$^{+0.3}_{-0.4}$ & 10.47 & \nodata & 10.61 & 10.84 & BD(P) & 51 & 35$^{+13}_{-10}$ & 61$^{+20}_{-20}$ & GRAVITY & G20 \\
MCG +04-22-042 & 7.59$^{+0.42}_{-0.28}$ & 9.92 & 10.09 & 10.55 & 10.70 & BDB(P) & 58 & 11$^{+6}_{-5}$ & 14$^{+7}_{-5}$ & CARAMEL & V22 \\
Mrk 110 & 7.17$^{+0.67}_{-0.26}$ & 9.57 & \nodata & \nodata & 9.57 & B(C) & \nodata & 20$^{+10}_{-11}$ & 27$^{+16}_{-13}$ & CARAMEL$^1$ & V22 \\
Mrk 1239 & 7.47$^{+0.15}_{-0.92}$ & 9.25 & \nodata & 9.55 & 9.73 & BD(P) & 43 & 11$^{+6}_{-3}$ & 42$^{+18}_{-15}$ & GRAVITY & G24 \\
Mrk 141 & 7.46$^{+0.15}_{-0.21}$ & 9.85 & 10.12 & 10.54 & 10.69 & BDB(P) & 40 & 26$^{+6}_{-4}$ & 15$^{+4}_{-2}$ & CARAMEL & W18 \\
NGC 3227 & 7.04$^{+0.11}_{-0.11}$ & 8.55 & \nodata & 9.31 & 9.38 & BD(C) & 67 & 33$^{+14}_{-9}$ & 65$^{+18}_{-12}$ & CARAMEL & B23a \\
Mrk 142 & 6.23$^{+0.3}_{-0.3}$ & 9.35 & 9.76 & 9.69 & 10.03 & BDB(P) & 36 & 41$^{+21}_{-11}$ & 30$^{+14}_{-12}$ & CARAMEL$^1$ & L18 \\
NGC 3516 & 7.61$^{+0.3}_{-0.7}$ & 9.59 & 9.87 & 9.77 & 10.13 & BDB(P) & 35 & \nodata & \nodata & cRM & D18$^{\ast\ast}$ \\
SBS 1116+583A & 6.99$^{+0.32}_{-0.25}$ & 8.85 & 9.15 & 9.65 & 9.78 & BDB(P) & 29 & 18$^{+8}_{-6}$ & 22$^{+11}_{-8}$ & CARAMEL & P14 \\
Arp 151 & 6.62$^{+0.1}_{-0.13}$ & 9.28 & \nodata & 9.24 & 9.56 & BD(C) & 71 & 25$^{+3}_{-3}$ & 26$^{+4}_{-4}$ & CARAMEL & P14 \\
NGC 3783 & 7.45$^{+0.07}_{-0.1}$ & 8.99 & 9.28 & 9.99 & 10.13 & BDB(P) & 29 & 18$^{+5}_{-6}$ & 35$^{+6}_{-10}$ & CARAMEL & B21b \\
Mrk 1310 & 7.42$^{+0.26}_{-0.27}$ & 9.17 & \nodata & 9.55 & 9.70 & BD(C) & 44 & 7$^{+5}_{-2}$ & 9$^{+4}_{-2}$ & CARAMEL & P14 \\
NGC 4151 & 7.22$^{+0.11}_{-0.1}$ & 9.17 & \nodata & 9.12 & 9.45 & BD(C) & 30 & 58$^{+10}_{-8}$ & 57$^{+14}_{-16}$ & CARAMEL & B22 \\
PG 1211+143 & 8.07$^{+0.11}_{-0.15}$ & 10.13 & \nodata & \nodata & 10.13 & B(C) & \nodata & \nodata & \nodata & cRM & K00$^{\ast\ast}$ \\
Mrk 50 & 7.51$^{+0.06}_{-0.07}$ & 9.71 & \nodata & 9.82 & 10.07 & BD(C) & 40 & 20$^{+6}_{-5}$ & 14$^{+5}_{-4}$ & CARAMEL & W18 \\
NGC 4593 & 6.65$^{+0.27}_{-0.15}$ & 9.52 & 9.90 & 8.33 & 9.91 & BDB(P) & 64 & 32$^{+19}_{-10}$ & 43$^{+22}_{-19}$ & CARAMEL & W18 \\
PG 1310-108 & 6.48$^{+0.21}_{-0.18}$ & 9.18 & 9.41 & 9.91 & 10.05 & BDB(P) & 24 & 44$^{+35}_{-13}$ & 58$^{+25}_{-16}$ & CARAMEL & W18 \\
RBS 1303 & 6.79$^{+0.19}_{-0.11}$ & 9.90 & 10.30 & 10.62 & 10.82 & BDB(P) & 55 & 29$^{+8}_{-9}$ & 34$^{+9}_{-10}$ & CARAMEL & V22 \\
IC 4329A & 7.64$^{+0.53}_{-0.25}$ & 8.76 & \nodata & 9.60 & 9.66 & BD(C) & 26 & 40$^{+27}_{-18}$ & 69$^{+15}_{-30}$ & CARAMEL & B23b \\
Mrk 279 & 7.58$^{+0.08}_{-0.08}$ & 10.35 & \nodata & 10.24 & 10.60 & BD(C) & 52 & 29$^{+3}_{-3}$ & 41$^{+4}_{-4}$ & CARAMEL & W18 \\
NGC 5548 & 7.54$^{+0.34}_{-0.24}$ & 10.11 & \nodata & 9.82 & 10.29 & BD(C) & 30 & 47$^{+13}_{-16}$ & 39$^{+14}_{-14}$ & CARAMEL$^2$ & W20 \\
PG 1426+015 & 9.02$^{+0.11}_{-0.15}$ & 10.48 & \nodata & \nodata & 10.48 & B(C) & \nodata & \nodata & \nodata & cRM & K00$^{\ast\ast}$ \\
Mrk 841 & 7.62$^{+0.5}_{-0.3}$ & 9.87 & 9.99 & 10.08 & 10.35 & BDB(P) & 19 & 30$^{+11}_{-15}$ & 41$^{+11}_{-11}$ & CARAMEL & V22 \\
Mrk 1392 & 8.16$^{+0.11}_{-0.13}$ & 9.92 & 10.24 & 10.67 & 10.83 & BDB(P) & 61 & 26$^{+3}_{-3}$ & 41$^{+5}_{-5}$ & CARAMEL & V22 \\
Mrk 1511 & 7.11$^{+0.2}_{-0.17}$ & 9.06 & 9.98 & 10.47 & 10.62 & BDB(P) & 20 & 19$^{+6}_{-5}$ & 36$^{+9}_{-10}$ & CARAMEL & W18 \\
PG 1617+175 & 7.69$^{+0.21}_{-0.38}$ & 10.27 & \nodata & \nodata & 10.27 & B(C) & \nodata & \nodata & \nodata & cRM & H21$^{\ast\ast}$ \\
NPM 1G+27.0587 & 7.64$^{+0.4}_{-0.36}$ & 10.10 & \nodata & 11.00 & 11.05 & BD(P) & 39 & 19$^{+11}_{-8}$ & 18$^{+11}_{-9}$ & CARAMEL & V22 \\
Zw 229-015 & 6.94$^{+0.14}_{-0.14}$ & 9.52 & 10.02 & 9.85 & 10.27 & BDB(P) & 49 & 33$^{+6}_{-5}$ & 34$^{+6}_{-6}$ & CARAMEL & W18 \\
NGC 6814 & 6.42$^{+0.24}_{-0.16}$ & 8.37 & 8.91 & 9.52 & 9.72 & BDB(P) & 24 & 49$^{+20}_{-22}$ & 50$^{+22}_{-19}$ & CARAMEL & P14 \\
RXJ 2044.0+2833 & 7.09$^{+0.17}_{-0.17}$ & 9.40 & 10.02 & 10.21 & 10.57 & BDB(P) & 47 & 42$^{+10}_{-8}$ & 51$^{+15}_{-12}$ & CARAMEL & V22 \\
Mrk 509 & 8.0$^{+0.06}_{-0.23}$ & 10.15 & \nodata & 10.04 & 10.40 & BD(C) & 40 & 69$^{+6}_{-12}$ & 64$^{+11}_{-9}$ & GRAVITY & G24 \\
PG 2130+099 & 6.92$^{+0.24}_{-0.23}$ & 9.35 & \nodata & 9.91 & 10.02 & BD(C) & 54 & 30$^{+11}_{-10}$ & 33$^{+12}_{-12}$ & CARAMEL & G17 \\
PG 2209+184 & 7.53$^{+0.19}_{-0.2}$ & 10.62 & \nodata & 10.23 & 10.77 & BD(C) & 30 & 30$^{+9}_{-7}$ & 29$^{+11}_{-8}$ & CARAMEL & V22 \\
RBS 1917 & 7.04$^{+0.23}_{-0.35}$ & 9.31 & \nodata & 9.91 & 10.01 & BD(P) & 26 & 20$^{+10}_{-4}$ & 25$^{+9}_{-8}$ & CARAMEL & V22 \\
NGC 7469 & 7.18$^{+0.05}_{-0.09}$ & 9.71 & 10.19 & 9.84 & 10.37 & BDB(P) & 37 & \nodata & \nodata & cRM & L21$^{\ast\ast}$ \\
\enddata
                   \tablecomments{
Col. (1): AGN Name.		      	   	  
Col. (2): Logarithm of \mbh~(solar units), compiled from literature (see column 12 for references).
Col. (3): Logarithm of spheroid 
luminosity (solar units)
(uncertainty of 0.04 dex).
Luminosities given in columns (3-6) are either in $I$ band
or $V$ band,
depending on filter used (see Table~\ref{tab:hst}).
Col. (4): Logarithm of spheroid+bar 
luminosity (solar units)
(uncertainty of 0.04 dex).
Col. (5): Logarithm of disk 
luminosity (solar units)
(uncertainty of 0.04 dex).
Col. (6): Logarithm of host 
luminosity (solar units)
(uncertainty of 0.04 dex).
Col. (7): Host-galaxy fit (B: spheroid only, BD: spheroid+disk, BDB:
spheroid+disk+bar). In parentheses: Spheroid component: C = classical
bulge; P = pseudo-bulge.
Col. (8): Host inclination based on disk axis ratio 
(Eq.~2) (uncertainty of $\pm$5 $\degr$).
Col. (9): BLR inclination from BLR modeling. 
Col. (10): BLR opening angle from BLR modeling. 
Col. (11): Sample based on \mbh~measurement.
$^1$: For three objects (Ark\,120, Mrk\,110, and Mrk\,142),
BLR models fit the data only with moderate quality \citep{Villafana:2022}.
$^2$: For NGC\,5548, dynamical BLR modeling was performed for two different data sets, set apart by 6 years \citep{Pancoast:2014, Williams:2020}. Results used here are taken from \citet{Williams:2020}, based on H$\beta$ versus V-band, for consistency.
Col. (12): References for \mbh~measurements and BLR modeling (where applicable).
$^{\ast}$: For NGC\,4593, the HST FOV is too small to cover the entire disk and thus, the disk fit is unreliable.  
$^{\ast\ast}$: Denotes objects for which \mbh~was re-calculated here, assuming an average
log $f$ = 0.65.
B21b: \citet{Bentz:2021b};
B23a: \citet{Bentz:2023a};
B23b: \citet{Bentz:2023b};
D15: \citet{Du:2015};
D18: \citet{DeRosa:2018};
F17: \citet{Fausnaugh:2017};
G17: \citet{Grier:2017};
G20: \citet{GravityCollaboration:2020};
G21b: \citet{GravityCollaboration:2021b};
G24: \citet{GravityCollaboration:2024};
H21: \citet{Hu:2021};
K00: \citet{Kaspi:2000};
L18: \citet{Li:2018};
L21: \citet{Lu:2021}
M21: \citet{Mandal:2021};
P14: \citet{Pancoast:2014}
U22: \citet{U:2022};
V22: \citet{Villafana:2022};
W18: \citet{Williams:2018};
W20: \citet{Williams:2020}.
}
            \label{tab:results}
\end{deluxetable*}

\subsection{\mbh-Spheroid Luminosity Relation}
\begin{figure*}[t]
  \centering
  \includegraphics[]{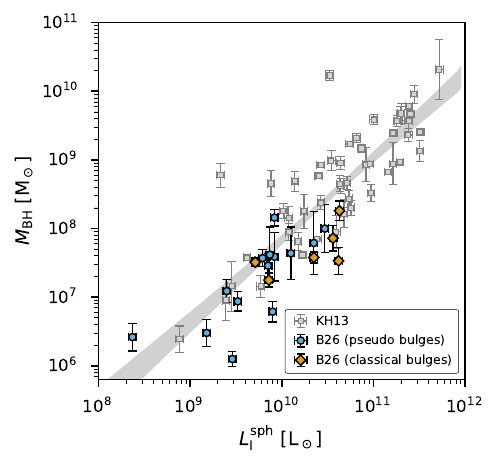}
  \includegraphics[]{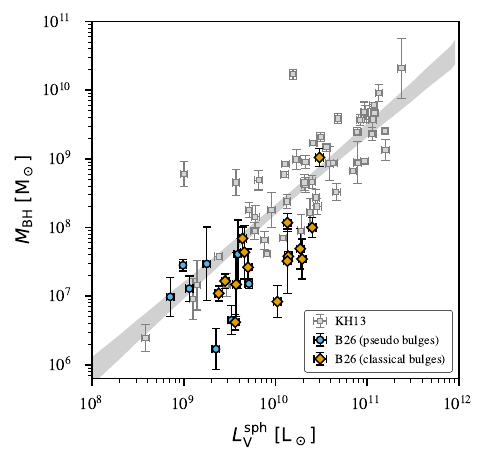}
      \caption{
        \emph{\mbh--Spheroid Luminosity Relation.}
In the left panel, the relation between \mbh~and spheroid luminosity
in the $I$-band is shown, in the right panel, the same but in the $V$-band.
Grey data points are quiescent galaxies 
\citep[ellipticals or spirals/S0 with classical bulges;][]{Kormendy_Ho:2013};
colored data points are our sample (classical bulges in yellow, pseudo-bulges in blue). 
The fitted relation is shown as a shaded gray stripe
corresponding to the  68\% (1-sigma) confidence 
interval of the linear regression.
Since our sample consists of local AGNs with directly measured \mbh~through dynamical modeling which is free of assumptions
of the virial factor, we can show that
AGNs follow the same scaling relations as those
of quiescent galaxies.}
            \label{fig:relation}
\end{figure*}

Figure~\ref{fig:relation} shows the resulting relation between
\mbh~and spheroid luminosity, split
 into $I$-band and $V$-band luminosity,
according to the HST filter used.
Overall, our AGN sample with directly measured BH masses follows
the same \mbh-spheroid-luminosity relation as quiescent galaxies,
and naturally extends the relation to lower masses, for these (predominantly) spiral host galaxies.
We fit the relations using a linear regression, taking into account
uncertainties \citep[$\texttt{LinMix}$;][]{Kelly:2007}
\begin{equation}
\label{eq:relation}
{\rm log} \left( \frac{M_{\rm BH}}{M_\odot}\right) = \alpha + \beta\, \log  X
\end{equation}
with $X$ either $I$-band or $V$-band luminosities
({$L^{\rm sph}_{I}  / 10^{10} L_\odot$} or  {$L^{\rm sph}_{V}  / 10^{10} L_\odot$}).
Overall,
the scaling relations between \mbh\ and spheroid luminosity agree 
for AGNs and quiescent galaxies within the uncertainties
(Table~\ref{tab:linmix}).
This further strengthens the conclusion reached by our previous paper
based on the same sample \citep{Winkel:2025a},
where we demonstrate, for the first time, that AGNs follow
the same \mbh--\sigstar~and \mbh--\mdyn~relations as quiescent galaxies,
regardless of host-galaxy morphology, when considering the spheroid component.
Our sample has the most robust \mbh~determined
free of assumptions on the virial factor.
Thus, this is an important independent
confirmation of the same underlying scaling relations between AGNs and
quiescent galaxies, implying that
AGNs can indeed be simply considered phases in the evolution of galaxies during which the
SMBH is actively growing through accretion.

The majority of AGN host galaxies in our sample are spiral
galaxies with roughly half of them having pseudo-bulges
\citep[see also e.g.,][for a similar fraction in a sample of low-redshift PG quasars]{Zhao:2021}.
Unlike merger-induced classical bulges,
pseudo-bulges are thought to have formed through secular
evolution via dissipative processes instead of galaxy mergers
\citep[e.g.,][]{Courteau:1996,Korista_Goad:2004},
experiencing recent or ongoing star formation
\citep[][]{Zhao:2021}.

In the past, pseudo-bulges have been found to preferentially lie off the scaling relations of ellipticals and classical bulges, especially in quiescent galaxies (\citealt{Kormendy_Ho:2013}; see, e.g., \citealt{Rios-Lopez:2025} for a recent compilation).
In the sample studied here, they do not form particular outliers in the relations, in agreement with other studies based on local type-1 AGNs \citep[e.g.,][]{Bennert:2021}.
It indicates that major mergers alone are not the only route to creating scaling relations,
adding to predictions from purely hierarchical growth 
models, which posit a non-causal origin for the relations
\citep[e.g.,][]{Jahnke:2011}. Instead, AGN feedback or other mechanisms that result in fixed relative efficiencies of star formation and BH growth could play an essential role in the co-evolution of SMBHs and their hosts.

When considering the total host-galaxy luminosity,
instead of just
the spheroid component,
we find their BHs to be residing in over-luminous hosts.
In other words, \mbh\ scales more closely with spheroid luminosity than host-galaxy luminosity, in agreement with other studies
\citep{Kormendy_Ho:2013, Kormendy:2011}, and similar to the tighter correlations of \mbh~and spheroid stellar velocity dispersion and stellar mass.
Finding over-luminous host galaxies of these local AGNs
suggests enhanced star formation (SF)
in these AGN host galaxies. However, in the literature, 
this topic is discussed quite controversially,
with studies reporting both 
enhanced as well as suppressed SF
in AGN host galaxies, or that SF is independent of AGN activity, with stellar mass
being the underlying driving factor
\citep[see, e.g.,][for a detailed
discussion]{Suh:2019}.

\begin{deluxetable*}{lcccc}[t]
\tabletypesize{\small} 
\tablecaption{Fits to \mbh--$L_{\rm sph}$ Scaling Relations}
\tablehead{ 
    \colhead{$X$ in Relation}  & 
    \colhead{Sample fitted}  &   
    \colhead{$\alpha$}  &  
    \colhead{$\beta$}  & 
    \colhead{$\epsilon$}     }
\colnumbers
\startdata
{$L^{\rm sph}_{I}  / 10^{11} L_\odot$} & KH13  &  9.1\,$\pm$\,1.2  &  1.05\,$\pm$\,0.11  &  0.48\,$\pm$\,0.06\\
    & KH13+AGN &  9.0\,$\pm$\,1.0  &  1.20\,$\pm$\,0.09  &     0.51\,$\pm$\,0.05\\    
    \hline
    {$L^{\rm sph}_{V}  / 10^{11} L_\odot$} & KH13  & 9.5\,$\pm$\,1.2  &  1.08\,$\pm$\,0.11  &  0.49\,$\pm$\,0.06\\
    & KH13+AGN &  9.4\,$\pm$\,0.9  &  1.16\,$\pm$\,0.09  &  0.52\,$\pm$\,0.05\\
 \hline
\enddata
\tablecomments{
Scaling relations of the form ${\rm log}(M_{\rm BH}/M_\odot) = \alpha+ \beta \,{\rm log}X$.
All fits calculated in this paper.  
Col. (1):  $X$ used in the scaling relation.
Col. (2): Sample for which the \mbh--$L_{\rm sph}$ relation was fitted. 
KH13 is the comparison sample of 51 quiescent galaxies \citep{Kormendy_Ho:2013}. ``AGN'' refers to sample of local RM AGNs in this paper. 
Col. (3): Best-fit intercept of the \mbh--$L_{\rm sph}$ relation.
Col. (4): Best-fit slope.
Col. (5): Best-fit intrinsic scatter.
}                              
            \label{tab:linmix}
\end{deluxetable*}

\subsection{BLR Inclination and Dependency on Other Properties}
The BLR inclination derived by 
\texttt{CARAMEL} modeling is a tracer of the orientation of the central AGN engine. We here compare it with 
the host-galaxy disk inclination as well as
properties of the AGN (BLR opening angle, accretion disk inclination and radio jet inclination).

\subsubsection{BLR Inclination vs.\ Host Galaxy Disk Inclination}
A key open question relevant to SMBH fueling
is the relative orientation between the central
AGN (with the accretion disk size on the order of a few light days)
and its host galaxy (on the order of tens of kiloparsecs).
So far, indirect tracers of the AGN inclination or position angle have been used, such as radio jets,
narrow lines or dust emission,
assuming the standard
AGN unified model \citep{Antonucci:1993}.
Results have been mixed.
For example,
no evidence for alignment has been found between
the host galaxy stellar disks and
accretion disk megamasers
\citep[with the latter generally, but not always, being well aligned
with radio jets/lobes and ionization cones;][]{Greenhill:2009}.
Likewise, no alignment has been found between
the host-galaxy major
axis and the narrow-line region (NLR) traced by 
extended [\ion{O}{3}] emission
\citep[which has been found to align
with the radio jet axis;][]{Schmitt:2003,Fischer:2013}.
Previous studies using radio jets as tracers of the inner AGN orientation
have not found evidence for alignment  with the host galaxy either
\citep[see, e.g.,][and references
therein]{Kinney:2000,Schmitt:2002,Battye:2009}.
The most recent compilation, however, compares Very Long Baseline
Interferometry (VLBI) images of 5853 radio-loud AGNs (typically residing in
elliptical galaxies) with
Dark Energy Spectroscopic Instrument Legacy
Imaging Surveys data (DESI LS), finding
``a weak but significant alignment signal [...] between the
parsec-scale AGN jet and the kpc projected minor axis of the optical
host galaxy'' \citep{FernandezGil:2025}.
A prominent minor-axis alignment,
depending on galaxy shape and radio power was also found by
\citet[][]{Zheng:2024} for 3682 radio-loud AGNs.

For spiral galaxies, interpreting their (weak) radio emission is often
not as straightforward, as the radio emission may originate from 
supernovae, and thus trace the star-forming disk.
Moreover, weak jets may be bent as
they encounter the dense ISM in the disk.
Here, we are in the unique position to 
compare the BLR inclination and
host-galaxy disk inclination directly, for a sample of 34 spiral galaxies.
For 32 AGNs with dynamical modeling of RM data,
BLR size and geometry have been derived, including
measurements of BLR inclination and opening angle
(see Table~\ref{tab:results}, Column 12 for references).
For another three AGNs, GRAVITY modeling also includes
details on BLR geometry
\citep{GravityCollaboration:2020,
GravityCollaboration:2021b,
GravityCollaboration:2024}.
Out of these 35 galaxies, two (Mrk\,110, 3C\,120)
were classified as ellipticals and fit with
a spheroid-only component; we thus exclude them here.

For the host galaxy, we use the disk axis ratio ($q = b/a$) derived by \texttt{lenstronomy} to calculate inclination as follows
\begin{eqnarray}
\cos^2 i = (q^2 - q_0^2)/(1-q_0^2)
\end{eqnarray}
with $q_0$ = 0.2
\citep[following, e.g.,][]{deGrijs:1998,Ho:2011}.
We verified
visually, that this value is indeed a reliable measure of the disk
inclination, within the uncertainties
\citep[see discussion in][]{Winkel:2025a}.
Note that in \citet{Winkel:2025a}, $i = \arccos(b/a)$ was used, but the difference is $<1 \degree$ for most objects and $<3 \degree$ in all cases.
Figure~\ref{fig:inclination} (left panel) shows the comparison for 34 AGNs.

The parameter space is limited due to selection effects on both ends. 
The lowest disk inclination we derive from our fitting is $\sim$ 18$\degr$ ($q_{\rm disk} = 0.95$), which is likely a consequence of  constraining the disk to be more elliptical than the spheroid component
during the fitting process (see (8a) in Section~\ref{sec:sbp}).
At very high inclination,
obscuration makes it harder to detect Seyferts in spiral galaxies
\citep[see, e.g.,][]{Hunt:1999}.
Given that RM is based on broad-line AGNs, we would expect to preferentially
select BLRs that are seen more closely to face on (for disk-like BLRs).
According to the standard unified model,
for type-1 AGNs, the BLR is viewed directly, at inclinations of $\sim$0--30$\degr$,
while for type-2 AGNs, the BLR is obscured due to high inclination angles ($\sim$70--90$\degr$ \citep{Antonucci:1993}.
Indeed, all AGNs have BLR inclinations $<$70$\degr$, with the majority
between 10 and 40$\degr$ (average 29$\degr$).
However, for a discussion on the complexity of the viewing angle
of Seyfert types, see, e.g., \citet{Ramos_Padilla:2022} and references therein.

Regardless, the sample covers a large range in
both BLR and host-galaxy disk inclinations ($\sim$18--70\degr).
There is no statistically significant evidence of a
correlation between BLR and  host-galaxy disk inclination angles.
If at all, there is an indication of a weak anti-correlation
(standard Pearson correlation coefficient of $-0.16$).
A similar conclusion has previously been reached  by
\citet{Du:2025},  based on a much smaller sample of 8 objects
(which are also included in this study),
and calculating galaxy inclination angles from 2MASS $K_s$-band images.

Our result is seemingly contradictory to the recent findings of radio jets
being aligned with the host-galaxy minor axis
\citep{Zheng:2024, FernandezGil:2025}.
One reason for the different results may be the different host
galaxy morphologies.
The radio-loud AGNs in these studies are known to be typically hosted by elliptical
galaxies.
Our sample, however, as well as megamaser host galaxies are spiral galaxies.
Another study of 18 disk galaxies with extended double-lobed radio structures also found a misalignment \citep{Wu:2022}.
While  high-resolution simulations
show that misalignments between AGNs and host galaxies are
expected for both galaxy major mergers as well as for instabilities in isolated disks \citep{Hopkins:2012}, the observed differences in inner AGN (mis-)alignment with outer host galaxy have been interpreted differently by \citet{FernandezGil:2025}.
They argue that spiral galaxies,
predominantly formed through secular evolution and gas accretion from
cosmic filaments, fuel SMBHs through minor mergers and cosmic gas
accretion which occur isotropically, and thus result in a random orientation
between accretion disk and host-galaxy disk.
Major mergers, on the other hand, form a spheroid elongated in the
direction of the incoming companion, with tidal debris fueling the
SMBH along the same direction.
Our findings are in agreement with this interpretation.

\subsubsection{BLR Inclination vs.~BLR Opening Angle}
\begin{figure*}[t]
  \centering
    \includegraphics[]{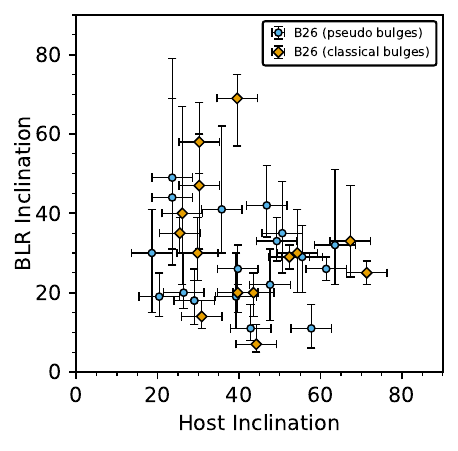}
    \includegraphics[]{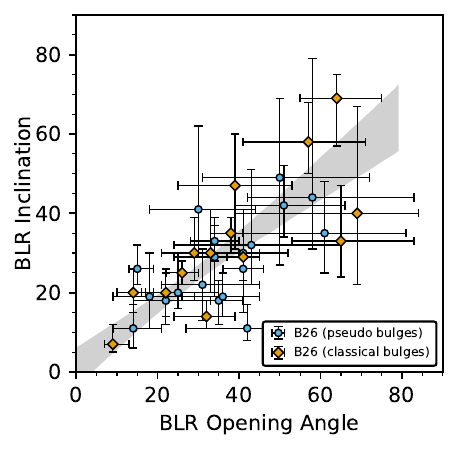}
    \caption{
        {\it Left:} \emph{BLR Inclination vs.~Host Inclination.}
        Comparison between inclination of the BLR, based on
        \texttt{CARAMEL} and GRAVITY modeling, and host-galaxy disk inclination, for our sample of 34 RM AGNs (classical bulges in yellow, pseudo-bulges in blue). 
        There is no correlation between the inclination
        of the central AGN and the large-scale host-galaxy disk;
        if at all, there is an indication of a weak anti-correlation
(Pearson correlation coefficient of $-0.16$).
        {\it Right:} \emph{BLR Inclination vs.~BLR Opening Angle.}
        Similar to the left panel, but now comparing 
 inclination and opening angle of the BLR, based on
        \texttt{CARAMEL} and GRAVITY modeling.
Both parameters are correlated (Pearson correlation coefficient of
0.71). The fitted relation is shown as a shaded gray stripe
corresponding to the  68\% (1-sigma) confidence 
interval of the linear regression.
The correlation is
driven by the absence of double-peaked lines in the sample.
In other words, in modeling the data, the opening angle of the BLR cannot be much smaller than the
inclination angle, while still reproducing the observed single-peaked lines.}
              \label{fig:inclination}
\end{figure*}

While the host-galaxy disk inclination angle can be relatively robustly determined
for these well-resolved, nearby galaxies imaged by HST,
we here address the question of possible degeneracies in the \texttt{CARAMEL} model.
BLR inclination is only one of the parameters used in \texttt{CARAMEL} to
describe BLR geometry; another important related one is the BLR
opening angle, which changes the geometry from a thin disk to a sphere.
In Figure~\ref{fig:inclination} (right panel), we show the distributions of
BLR inclination angles vs.~BLR opening angles.
Here, the Pearson correlation coefficient is 0.71, indicating that
both parameters are correlated.
\texttt{LinMix}
gives a best-fit linear regression of
$i_{\rm BLR}$ = (2$\pm$5)$\degr$ + (0.8$\pm$0.2) $o_{\rm BLR}$
with an intrinsic scatter of (5$\pm$2)$\degr$.
The relation between BLR inclination and opening angle was
noted and discussed already in \citet{Grier:2017}, based on a much smaller sample.
To understand this relation, it is important to remember what these parameters mean.
The smaller the inclination angle, the more face-on the BLR is seen;
the larger, the more edge-on. The smaller the opening angle, the more
disk-like is the BLR; the larger the more spherical
\citep[see, e.g., Figure 9 in][]{Pancoast:2014}.
For a given inclination angle, if the opening angle becomes too small
(i.e., disk-shaped), the emission line profile will show double peaked lines.
However, double-peaked emission lines are not seen in this sample,
so the opening angle cannot be much smaller than the inclination
angle.

On the other hand, as the opening angle becomes larger than the 
inclination angle, the transfer function becomes spread out in both
time lag and wavelength space. 
As \citet{Grier:2017} summarize: 
``If the data prefer
a more compact transfer function, it will thus force the opening
angle to be as small as possible while still producing a single-peaked
line profile.''
So the correlation between BLR inclination and opening angle
is not driven by a degeneracy in the \texttt{CARAMEL} model,
but rather by the physical requirements of the observed data (transfer function and line profiles).
\citet{Grier:2017} perform tests varying inclination and opening
angles, suggesting that the inclination angle is more robustly
determined than the opening angle.
Note that, naturally, the uncertainties on the derived inclination angles increase towards more
spherical BLRs (i.e., large opening angles), as the inclination cannot be
robustly determined for a spherical shape.

\subsubsection{BLR Inclination vs.~Accretion Disk and Jet Inclination}
Further support for the robustness of the BLR inclination angle
determined by \texttt{CARAMEL} comes from two independent approaches:
(1) using X-ray data to estimate the accretion-disk inclination;
(2) using radio data to determine radio jet inclination.

(1) For eight AGNs (also included in our sample), 
\citet{Du:2025} 
derive the accretion disk inclination angle based on
X-ray broadband reflection spectroscopy.
When compared to the 
BLR inclination angle (as derived from CARAMEL),
  they report a nearly linear relation (even though with
marginal significance, given their small sample).
The relation suggests that the rotational axes of BLR and accretion disk are aligned, supporting 
a physical connection between accretion disk and BLR,
e.g., through disk winds \citep{Peterson:2006, Czerny_Hryniewicz:2011}.
Taking an Occam's Razor approach, we interpret the nearly linear relation 
between BLR and accretion disk inclinations
as evidence for the robustness of the
the BLR
inclination angles determined by \texttt{CARAMEL}.

(2) While the majority of AGNs in the \texttt{CARAMEL} sample are radio-quiet, 
there are three objects for which 
 the radio jet inclination axis has been determined.
Two AGNs are radio-loud  (3C\,120 and
3C\,273) and one is a so-called radio-intermediate quasar
\citep[Mrk1501, also known as IIIZw\,2 or PG 0007+106;
thought to be relativistically boosted counterparts of radio-quiet
quasars; e.g.,][]{Falcke:1996}.
For 3C\,120, the radio jet inclination $i_{\rm
  radio~jet}=16\pm2 \degr$ \citep{Agudo:2012} is, within the uncertainties, identical to the BLR inclination
$i_{\rm BLR} =17.6^{+5.4}_{-3.3}$$\degr$~\citep[CARAMEL;][]{Grier:2017}.
The host is an elliptical galaxy with signs of interactions (one-armed tidal tail).
For 3C\,273, the radio jet inclination $i_{\rm radio~jet}=5.5\pm1.7 \degr$ \citep{Meyer:2016} also agrees well
with the BLR inclination  
$i_{\rm BLR} =5\pm1 \degr$~\citep[GRAVITY/SARM;][]{Li:2022}. 
This radio-loud AGN is hosted by an elliptical
galaxy that was not included in the HST decomposition presented here, due
to an insufficient PSF fit
\citep[however, see][for a decomposition]{Bentz:2009a}.
Lastly, for Mrk\,1501, $i_{\rm radio~jet} = 22.1$$\degr$ \citep{Liodakis:2018}
compares to $i_{\rm BLR} =20.5^{+5}_{-5.7}$$\degr$~\citep[CARAMEL;][]{Grier:2017},
with a host-galaxy disk inclination of 52$\degr$.
In all three cases, the radio jet inclination 
agrees well with the BLR inclination derived by the dynamical
modeling.

\begin{figure*}[t]
  \centering
    \includegraphics[]{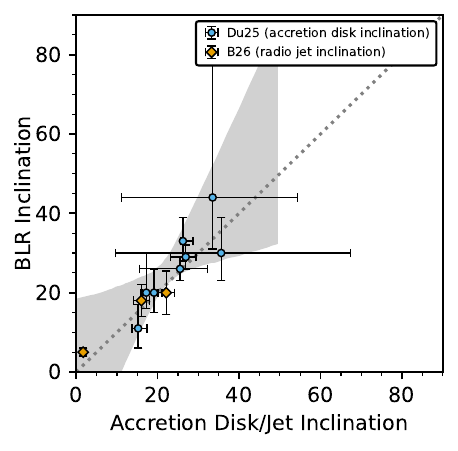}
    \caption{
        \emph{BLR Inclination vs.~Accretion Disk/Jet Inclination.}
        Comparison between  inclination of the BLR, based on
        \texttt{CARAMEL} modeling and accretion disk inclination 
        \citep{Du:2025} or radio-jet inclination (compiled in this paper).
Both parameters are correlated (Pearson correlation coefficient of
0.9). The fitted relation is shown as a shaded gray stripe
corresponding to the  68\% (1-sigma) confidence 
interval of the linear regression.
(The 1:1 dotted line is shown for comparison only.)}
              \label{fig:accretion_jet}
\end{figure*}

The relation between BLR inclination
and either (1) accretion disk inclination or (2) radio jet inclination is shown in Fig.~\ref{fig:accretion_jet}.
The Pearson correlation coefficient is 0.9, 
and \texttt{LinMix}
gives a best-fit linear regression of
$i_{\rm BLR}$ = ($-6$$\pm$25)$\degr$ + (1.4$\pm$1) $i_{\rm accretion\, disk/jet}$
with an intrinsic scatter of (5$\pm$5)$\degr$.

The close agreement between the BLR and accretion disk or jet inclination angles is remarkable, given the independent
derivations, at different wavelengths and scales.
These findings support the standard view of 
inner AGN structure, with 
the BLR, accretion disk
and jet aligned.
They are also further and independent evidence that the BLR inclination angle derived by \texttt{CARAMEL} can be considered robust.

\section{Summary}
\label{sec:summary}
In this paper, HST images are analyzed of a sample of 44 type-1 AGNs with the most robust BH mass measurements in the local universe.
Using surface-brightness profile fitting with state-of-the-art
programs, we derive AGN luminosity, as well as
radii and luminosities of spheroid, disk and bar (if present).
The spheroid and galaxy effective radii were used in \citet{Winkel:2025a}
to measure stellar velocity dispersion from integral-field
spectroscopy.
We  demonstrated -- for the first time -- that AGNs follow
the same \mbh--\sigstar~and \mbh--$M_{\rm sph, dyn}$ relations as
quiescent galaxies.
Here, we expand the study to other host-galaxy properties. Our results can be summarized as follows.

\begin{itemize}
\item[$\bullet$]{The majority of AGNs in our sample are hosted by
    spiral or S0 galaxies (91\%), with roughly half having a prominent
    bar. Also roughly half were determined to host pseudo-bulges. Major mergers are rare (5\%), with another $\sim$15--20\% showing signs of interactions or minor mergers.}
\item[$\bullet$]{Within the uncertainties, the \mbh--$L_{\rm sph}$
relation matches that of quiescent galaxies. Combined with the results in our earlier paper, 
for moderate luminosity AGNs in the local universe, all three scaling relations \mbh--\sigstar, \mbh--$M_{\rm sph,
  dyn}$ and \mbh--$L_{\rm sph}$  match those of quiescent
galaxies. Pseudo-bulges are not outliers.}
\item[$\bullet$]{For our sample of spiral galaxies, the BLR inclination  
and host-galaxy disk inclination angles are uncorrelated.
This is contrary to a recent studies of radio jets in
    elliptical galaxies for which an alignment between parsec-scale
    AGN jet and kpc projected minor axis of the optical host galaxy
    was found. The difference is likely due to the different nature and formation of their host galaxies.}
\item[$\bullet$]{BLR inclination and opening angle are positively
    related, driven by the need of the model to account for the
    single-peaked emission lines observed in the sample.}
\item[$\bullet$]{For three radio-loud AGNs in the sample, the radio
    jet inclination agrees well with the BLR inclination. For eight
    AGNs, the same is true between the accretion disk inclination and
    BLR inclination. These findings not only provide independent
    evidence that the derived BLR inclination angle is robust,
    they also support the standard
view of the inner AGN structure, with BLR, accretion disk and jet aligned.}
\end{itemize}

Our sample has the most accurate \mbh~measurements beyond the local universe and provides a fundamental local benchmark for studies of the evolution of massive black holes and their
host galaxies across cosmic time.

\begin{acknowledgements}
All HST data used in this paper can be found in MAST:
\href{https://doi.org/10.17909/2g2m-rx22}{doi:10.17909/2g2m-rx22}.
We thank the anonymous referee for their thoughtful comments that helped to improve the paper.
VNB thanks Eric Emsellem and Andrea Merloni for fruitful discussions, and Victoria Bollo and Pranav Kukreti for their help with the radio observations.
VNB gratefully acknowledges support through the European Southern
Observatory (ESO) Scientific Visitor Program. 
    This work is based on observations with the NASA/ESA Hubble Space Telescope obtained from the Data Archive at the Space Telescope Science Institute, which is operated by the Association of Universities for Research in Astronomy, Incorporated, under NASA contract NAS5-26555.
Financial support for Program number HST-GO 17103 (PI Bennert) and HST-AR
    17063 (PI Bennert) was provided through a grant from the STScI
    under NASA contract NAS5-26555.
 NW was supported by the German Science Foundation (DFG) under grant number HU\,1777/3–1.
MK was supported by the National Research Foundation of Korea (NRF) grant funded by the Korean government (MSIT) (No. RS-2024-00347548).
ChatGPT (version 4.0 and 5.0) was used for occasional minor help with Python coding (finding bugs, help with plotting, etc.) and minor text edits (typos and grammar).
\end{acknowledgements}

\facilities{Keck:II (KCWI), 
            VLT:Yepun (MUSE),
            VLT:Melipal (VIMOS),
            HST (ACS, WFCP2, WFC3)
    }

 \software{
        \texttt{AstroDrizzle}
        \citep{Fruchter:2010},
        \texttt{Astropy} \citep{AstropyCollaboration:2013,
          AstropyCollaboration:2018},
        \texttt{GALFIT} \citep{Peng:2002},
                          \texttt{galight} \citep{Ding:2020},
             \texttt{LA-Cosmic} \citep{vanDokkum:2001},
         \texttt{Lenstronomy} \citep{Birrer_Amara:2018},
         \texttt{LinMix} \citep{Kelly:2007},
             \texttt{psfr} \citep{Birrer:2019},
         \texttt{SciPy} \citep{SciPy:2020}
          }

\bibliographystyle{aasjournal}
\bibliography{references1,references2,references3}

\clearpage
\appendix

\section{Surface-Photometry Fitting}
\label{Appendix:Lenstronomy}
\begin{figure*}[p]
  \centering    
   \includegraphics[width=\textwidth,height=0.16\textheight,keepaspectratio]{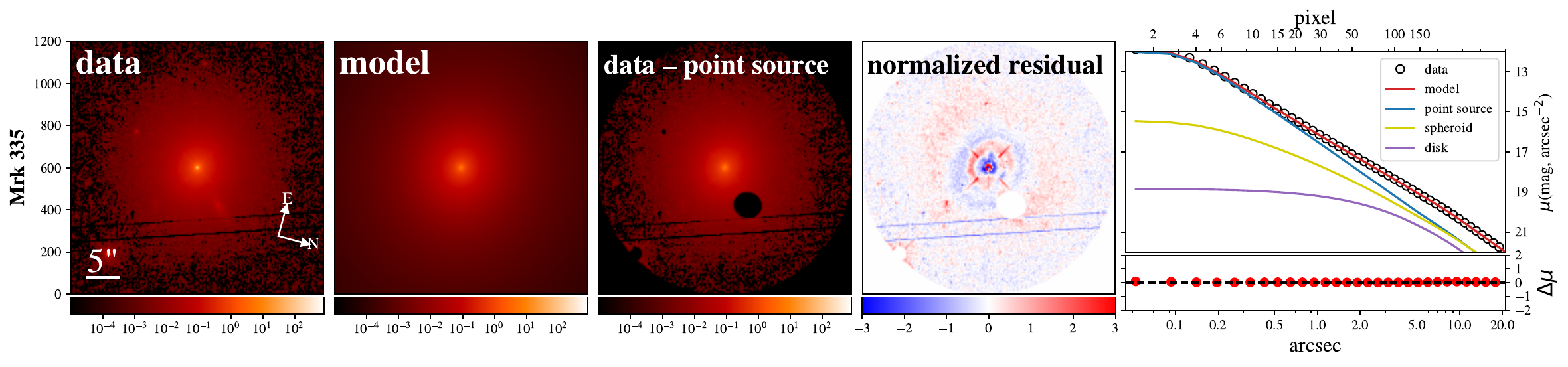}
   \includegraphics[width=\textwidth,height=0.16\textheight,keepaspectratio]{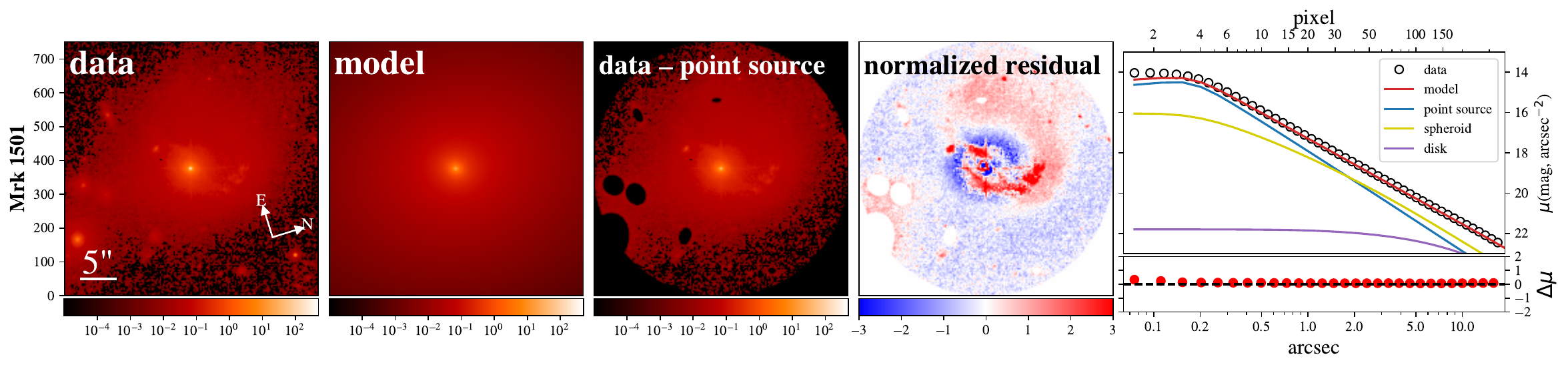}
      \includegraphics[width=\textwidth,height=0.16\textheight,keepaspectratio]{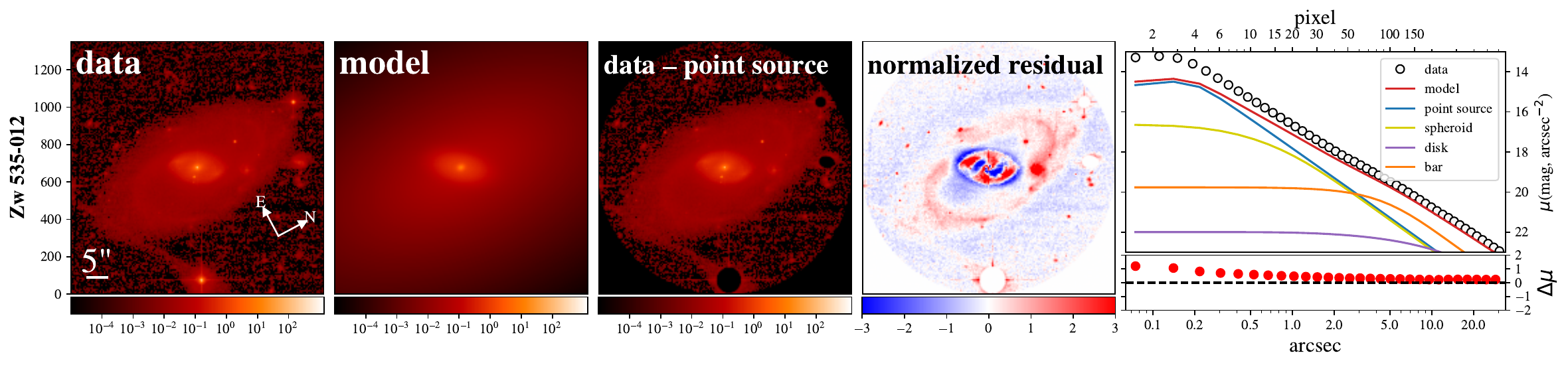}
      \includegraphics[width=\textwidth,height=0.16\textheight,keepaspectratio]{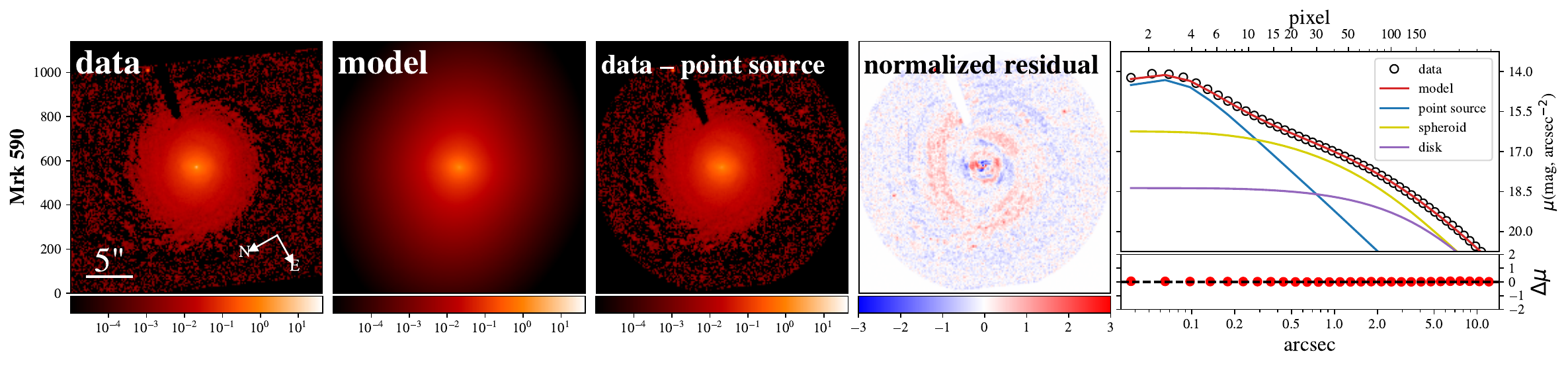}
      \includegraphics[width=\textwidth,height=0.16\textheight,keepaspectratio]{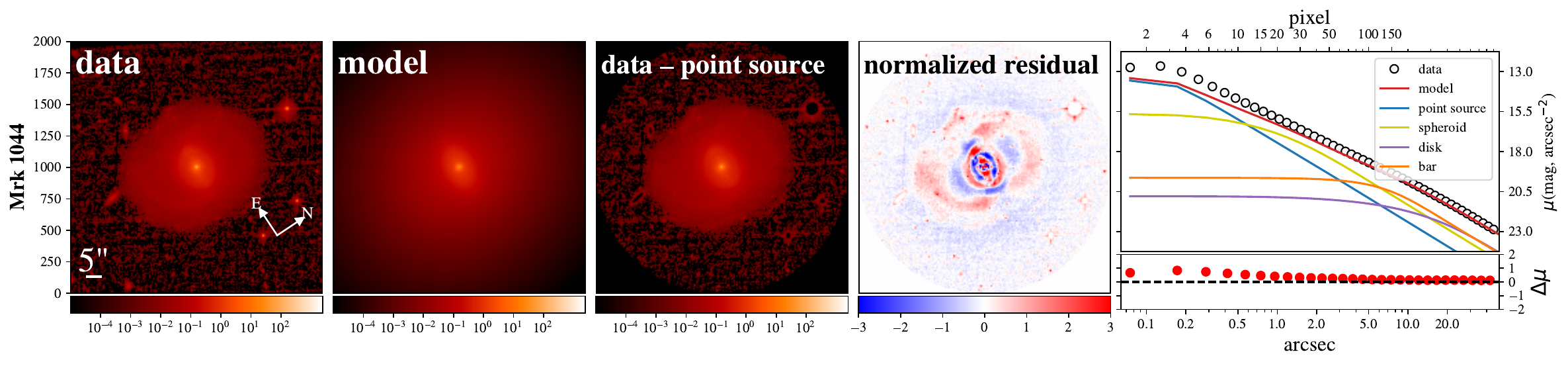} 
   \caption{
   \emph{Surface-Photometry Fits.}
From left to right: observed HST image (``data'', including 
scale bar of 5 arcseconds and North-East directions); best-fit \texttt{galight}/\texttt{lenstronomy} model (``model''); PSF-subtracted image (``data—point source''); residual image after subtraction of best-fit model
from data, divided by the noise level (``normalized residual'');
and surface-brightness profile 
(data = black circles, model = red line, PSF = blue line, 
spheroid = yellow line; if present: disk = purple line, bar = orange line).
The surface-brightness profile is shown for illustration only, as the fits were performed on the 2D image.
Surface-brightness values are given in the plane of the sky (total light within circular aperture; the x-axis is based on a circularized radius).
Note that all images
are displayed as observed with HST (see Table~\ref{tab:hst}) and fitted by \texttt{lenstronomy}.
Each row of images corresponds to one object (as labeled on the y-axis of
the leftmost panel). 
 }
              \label{fig:lenstronomy1}
\end{figure*}

\texttt{galight}/\texttt{lenstronomy} fitting results are summarized in Table~\ref{tab:SBPfittingresults}
 and fits are shown in Figure~\ref{fig:lenstronomy1} and the online figure set.
 Given the widespread use of \texttt{GALFIT} in the literature, \texttt{GALFIT} is
run for comparison, using the same background subtracted image, error
image, PSF and mask (if any). The fitting approach is similar to steps
(6--8) described in Section~\ref{sec:sbp}. Different starting parameters are used to ensure a true
global minimum has been reached. Residuals are visually inspected for
best fitting results. 
Overall, results agree and are robust for the spheroid-only fit in all
parameters,
as well as for the magnitudes in the spheroid-disk and spheroid-disk-bar
fits \citep[see also, e.g.,][]{Bennert:2021}. The largest scatter is found for the best-fit S{\'e}rsic index
for the spheroid component in spheroid-disk and spheroid-disk-bar
decompositions.
This cautions the sole reliance on S{\'e}rsic index to distinguish
between classical and pseudo-bulges.
Overall, the design of \texttt{lenstronomy} (semi-linear inversion and PSO)
makes results more robust.
Moreover, the automation significantly reduces the user-interaction
compared to 
\texttt{GALFIT} to ensure a true global minimum is reached in the fitting
process.

\begin{longrotatetable}
\begin{deluxetable*}{lccccccccccccccccc}
\tabletypesize{\footnotesize}
\tablecolumns{22}
  \tablecaption{Surface-Photometry Fitting Results}
\tablehead{
AGN Name & AGN & Spheroid & Disk
& Bar & $n_{\rm sph}$ &
$R_{\rm sph}$ & $R_{\rm sph}$ & PA$_{\rm sph}$ & $q_{\rm sph}$ &
$R_{\rm disk}$ & $R_{\rm disk}$ & PA$_{\rm disk}$ & $q_{\rm disk}$ & 
$R_{\rm bar}$ & $R_{\rm bar}$ & PA$_{\rm bar}$ & $q_{\rm bar}$ \\
& (mag) & (mag) & (mag) & (mag) 
&  & ($\prime\prime$) & (kpc) & ($\degr$) & & 
($\prime\prime$) & (kpc) & ($\degr$) & &
($\prime\prime$) & (kpc) & ($\degr$) \\
(1) & (2) & (3)  & (4) & (5) & (6) & (7) & (8) & (9)  & (10) & (11) &
(12) & (13)  & (14) & (15) & (16)  & (17) & (18)}
\startdata
Mrk 335 & 15.2 & 15.1 & 15.7 & \nodata & 4.4 & 2.32 & 1.24 & 270 & 0.91 & 2.9 & 1.56 & 265 & 0.91 & \nodata & \nodata & \nodata & \nodata \\ 
Mrk 1501 & 16.6 & 16.1 & 16.3 & \nodata & 5.0 & 1.29 & 2.17 & 141 & 0.89 & 9.75 & 16.39 & 176 & 0.74 & \nodata & \nodata & \nodata & \nodata \\ 
Zw 535-012 & 16.5 & 16.6 & 15.6 & 15.6 & 1.1 & 0.58 & 0.56 & 170 & 0.8 & 16.81 & 16.19 & 175 & 0.53 & 4.3 & 4.14 & 140 & 0.53 \\ 
Mrk 590 & 18.0 & 15.2 & 15.2 & \nodata & 1.5 & 1.42 & 0.77 & 322 & 0.76 & 3.2 & 1.73 & 50 & 0.76 & \nodata & \nodata & \nodata & \nodata \\ 
Mrk 1044 & 16.2 & 15.1 & 14.6 & 14.6 & 1.2 & 0.77 & 0.26 & 162 & 0.95 & 12.08 & 4.09 & 185 & 0.88 & 5.81 & 1.97 & 84 & 0.63 \\ 
Mrk 1048 & 19.0 & 15.0 & 13.8 & \nodata & 1.0 & 2.68 & 2.33 & 120 & 0.69 & 15.42 & 13.4 & 80 & 0.69 & \nodata & \nodata & \nodata & \nodata \\ 
3C 120 & 15.0 & 16.2 & \nodata & \nodata & 5.0 & 2.87 & 1.95 & 301 & 0.97 & \nodata & \nodata & \nodata & \nodata & \nodata & \nodata & \nodata & \nodata \\ 
Ark 120 & 14.2 & 13.7 & 14.7 & \nodata & 3.7 & 2.96 & 1.99 & 358 & 0.87 & 11.6 & 7.82 & 21 & 0.86 & \nodata & \nodata & \nodata & \nodata \\ 
NGC 2617 & 16.2 & 15.8 & 13.3 & \nodata & 1.7 & 1.24 & 0.37 & 20 & 0.95 & 12.44 & 3.72 & 106 & 0.95 & \nodata & \nodata & \nodata & \nodata \\ 
IRAS 09149-6206 & 13.0 & 15.4 & 15.0 & \nodata & 1.0 & 2.44 & 2.8 & 47 & 0.84 & 9.9 & 11.34 & 70 & 0.65 & \nodata & \nodata & \nodata & \nodata \\ 
MCG +04-22-042 & 16.9 & 15.5 & 14.0 & 16.6 & 1.3 & 0.92 & 0.63 & 351 & 0.78 & 11.67 & 7.98 & 353 & 0.56 & 7.29 & 4.98 & 338 & 0.2 \\ 
Mrk 110 & 16.2 & 16.6 & \nodata & \nodata & 1.8 & 1.51 & 1.1 & 280 & 0.96 & \nodata & \nodata & \nodata & \nodata & \nodata & \nodata & \nodata & \nodata \\ 
Mrk 1239 & 15.0 & 15.8 & 15.1 & \nodata & 1.0 & 0.99 & 0.41 & 151 & 0.75 & 4.85 & 1.99 & 154 & 0.75 & \nodata & \nodata & \nodata & \nodata \\ 
Mrk 141 & 18.5 & 16.2 & 14.5 & 16.4 & 1.0 & 0.37 & 0.31 & 168 & 0.81 & 5.99 & 5.09 & 139 & 0.78 & 3.39 & 2.88 & 149 & 0.31 \\ 
NGC 3227 & 14.6 & 14.2 & 12.3 & \nodata & 2.5 & 1.79 & 0.14 & 167 & 0.64 & 30.83 & 2.47 & 152 & 0.43 & \nodata & \nodata & \nodata & \nodata \\ 
Mrk 142 & 16.4 & 17.7 & 16.9 & 17.2 & 1.0 & 0.41 & 0.37 & 17 & 0.83 & 7.66 & 6.94 & 19 & 0.82 & 3.48 & 3.15 & 41 & 0.46 \\ 
NGC 3516 & 15.7 & 13.5 & 13.1 & 13.8 & 1.1 & 1.96 & 0.37 & 234 & 0.83 & 16.3 & 3.05 & 210 & 0.83 & 7.9 & 1.48 & 344 & 0.57 \\ 
SBS 1116+583A & 18.0 & 17.9 & 15.9 & 18.0 & 1.0 & 0.59 & 0.34 & 68 & 0.88 & 4.62 & 2.67 & 64 & 0.88 & 2.95 & 1.7 & 70 & 0.31 \\ 
Arp 151 & 16.9 & 16.2 & 16.3 & \nodata & 2.8 & 1.21 & 0.53 & 155 & 0.76 & 5.33 & 2.31 & 157 & 0.37 & \nodata & \nodata & \nodata & \nodata \\ 
NGC 3783 & 14.6 & 15.2 & 12.7 & 15.7 & 1.0 & 2.04 & 0.42 & 293 & 0.88 & 15.15 & 3.12 & 160 & 0.88 & 12.62 & 2.6 & 160 & 0.19 \\ 
Mrk 1310 & 17.3 & 16.3 & 15.4 & \nodata & 4.8 & 4.2 & 1.71 & 308 & 0.87 & 4.2 & 1.71 & 320 & 0.73 & \nodata & \nodata & \nodata & \nodata \\ 
NGC 4151 & 13.0 & 12.4 & 12.6 & \nodata & 4.4 & 6.18 & 0.44 & 59 & 0.87 & 11.92 & 0.85 & 336 & 0.87 & \nodata & \nodata & \nodata & \nodata \\ 
PG 1211+143 & 14.7 & 17.0 & \nodata & \nodata & 5.0 & 0.15 & 0.24 & 288 & 0.59 & \nodata & \nodata & \nodata & \nodata & \nodata & \nodata & \nodata & \nodata \\ 
Mrk 50 & 16.5 & 15.3 & 15.0 & \nodata & 5.0 & 4.05 & 1.98 & 170 & 0.8 & 4.16 & 2.03 & 173 & 0.78 & \nodata & \nodata & \nodata & \nodata \\ 
NGC 4593 & 16.3 & 13.6 & 16.6 & 13.3 & 1.5 & 6.21 & 1.1 & 97 & 0.69 & 39.97 & 7.07 & 352 & 0.48 & 32.3 & 5.72 & 58 & 0.33 \\ 
PG 1310-108 & 15.8 & 17.5 & 15.6 & 18.0 & 1.0 & 0.4 & 0.28 & 75 & 0.92 & 3.68 & 2.59 & 77 & 0.92 & 1.25 & 0.88 & 113 & 0.39 \\ 
RBS 1303 & 15.1 & 16.1 & 14.3 & 15.8 & 1.1 & 0.94 & 0.8 & 285 & 0.68 & 10.39 & 8.85 & 288 & 0.59 & 4.01 & 3.42 & 163 & 0.45 \\ 
IC 4329A & 14.7 & 16.7 & 14.6 & \nodata & 5.0 & 1.97 & 0.63 & 130 & 0.9 & 3.13 & 0.99 & 240 & 0.9 & \nodata & \nodata & \nodata & \nodata \\ 
Mrk 279 & 15.0 & 14.3 & 14.6 & \nodata & 2.1 & 2.44 & 1.54 & 25 & 0.63 & 8.67 & 5.46 & 65 & 0.63 & \nodata & \nodata & \nodata & \nodata \\ 
NGC 5548 & 15.0 & 13.6 & 14.3 & \nodata & 3.4 & 7.85 & 2.68 & 86 & 0.96 & 15.16 & 5.18 & 118 & 0.87 & \nodata & \nodata & \nodata & \nodata \\ 
PG 1426+015 & 15.7 & 16.4 & \nodata & \nodata & 1.4 & 1.99 & 3.3 & 53 & 0.69 & \nodata & \nodata & \nodata & \nodata & \nodata & \nodata & \nodata & \nodata \\ 
Mrk 841 & 14.9 & 15.9 & 15.4 & 17.4 & 1.8 & 1.27 & 0.95 & 104 & 0.95 & 6.07 & 4.53 & 137 & 0.95 & 2.53 & 1.89 & 165 & 0.52 \\ 
Mrk 1392 & 16.2 & 15.7 & 13.8 & 15.8 & 1.4 & 0.74 & 0.54 & 220 & 0.76 & 14.04 & 10.34 & 228 & 0.51 & 4.21 & 3.1 & 155 & 0.47 \\ 
Mrk 1511 & 16.6 & 17.8 & 14.3 & 15.7 & 1.2 & 0.43 & 0.3 & 136 & 0.94 & 12.61 & 8.79 & 21 & 0.94 & 9.49 & 6.62 & 161 & 0.35 \\ 
PG 1617+175 & 16.1 & 17.5 & \nodata & \nodata & 5.0 & 1.19 & 2.51 & 40 & 0.87 & \nodata & \nodata & \nodata & \nodata & \nodata & \nodata & \nodata & \nodata \\ 
NPM 1G+27.0587 & 15.8 & 16.5 & 14.2 & \nodata & 1.0 & 0.61 & 0.76 & 204 & 0.89 & 6.61 & 8.16 & 259 & 0.78 & \nodata & \nodata & \nodata & \nodata \\ 
Zw 229-015 & 17.7 & 16.1 & 15.3 & 15.5 & 1.3 & 0.75 & 0.43 & 222 & 0.79 & 10.72 & 6.2 & 217 & 0.67 & 6.33 & 3.66 & 215 & 0.46 \\ 
NGC 6814 & 16.2 & 15.4 & 12.5 & 14.8 & 1.0 & 1.3 & 0.14 & 44 & 0.93 & 25.72 & 2.86 & 22 & 0.92 & 5.25 & 0.58 & 28 & 0.68 \\ 
RXJ 2044.0+2833 & 16.0 & 17.7 & 15.7 & 17.2 & 1.9 & 0.2 & 0.2 & 184 & 0.89 & 5.36 & 5.41 & 197 & 0.7 & 0.98 & 0.99 & 142 & 0.7 \\ 
Mrk 509 & 14.1 & 15.2 & 15.4 & \nodata & 5.0 & 1.84 & 1.31 & 70 & 0.78 & 2.61 & 1.86 & 82 & 0.78 & \nodata & \nodata & \nodata & \nodata \\ 
PG 2130+099 & 14.9 & 18.5 & 17.0 & \nodata & 3.8 & 0.34 & 0.43 & 57 & 0.61 & 2.17 & 2.75 & 51 & 0.61 & \nodata & \nodata & \nodata & \nodata \\ 
PG 2209+184 & 17.3 & 15.5 & 16.4 & \nodata & 5.0 & 2.85 & 3.93 & 99 & 0.95 & 3.03 & 4.19 & 93 & 0.87 & \nodata & \nodata & \nodata & \nodata \\ 
RBS 1917 & 16.1 & 18.6 & 17.1 & \nodata & 1.0 & 0.39 & 0.51 & 119 & 0.9 & 2.03 & 2.62 & 200 & 0.9 & \nodata & \nodata & \nodata & \nodata \\ 
NGC 7469 & 14.9 & 14.5 & 14.2 & 13.9 & 1.0 & 1.44 & 0.5 & 87 & 0.81 & 29.37 & 10.22 & 89 & 0.81 & 9.69 & 3.37 & 303 & 0.59 \\ 
\enddata
  \tablecomments{Surface-photometry fitting results using \texttt{galight}/\texttt{lenstronomy}
    on HST images. Note that the fitting was done on the original (randomly oriented) images, but all position angles given here are relative to north.
Col. (1): AGN Name.		      	   	  
Col. (2): Point-source (AGN) magnitude (uncertainty 0.1 mag).
Col. (3): Spheroid magnitude  (uncertainty 0.1 mag).
Col. (4): Disk magnitude (if present;  uncertainty 0.1 mag).
Col. (5): Bar magnitude (if present;  uncertainty 0.1 mag).
Col. (6): Spheroid S{\'e}rsic index $n$  (5\% uncertainty).
Col. (7): Spheroid radius in arcseconds (10\% uncertainty).
Col. (8): Spheroid radius in kpc. 
Col. (9): Spheroid position angle east of north in degrees (1 $\degr$ uncertainty).
Col. (10): Spheroid axis ratio $q$ (=$b/a$) (0.05 uncertainty).
Col. (11): Disk radius in arcseconds (10\% uncertainty).
Col. (12): Disk radius in kpc.
Col. (13): Disk position angle east of north in degrees (1 $\degr$ uncertainty).
Col. (14): Disk axis ratio $q$ (=$b/a$) (0.05  uncertainty).
Col. (15): Bar radius in arcseconds (10\% uncertainty).
Col. (16): Bar radius in kpc. 
Col. (17): Bar position angle east of north in degrees (1 $\degr$ uncertainty).
Col. (18): Bar axis ratio $q$ (=$b/a$) (0.05  uncertainty).
    $^{\ast}$: For NGC\,4593, the HST FOV is too small to cover the entire disk and thus, the disk fit is unreliable. 
}
  \label{tab:SBPfittingresults}
\end{deluxetable*}
\end{longrotatetable}

\begin{figure*}[p]
  \centering
   \includegraphics[width=\textwidth,height=0.16\textheight,keepaspectratio]{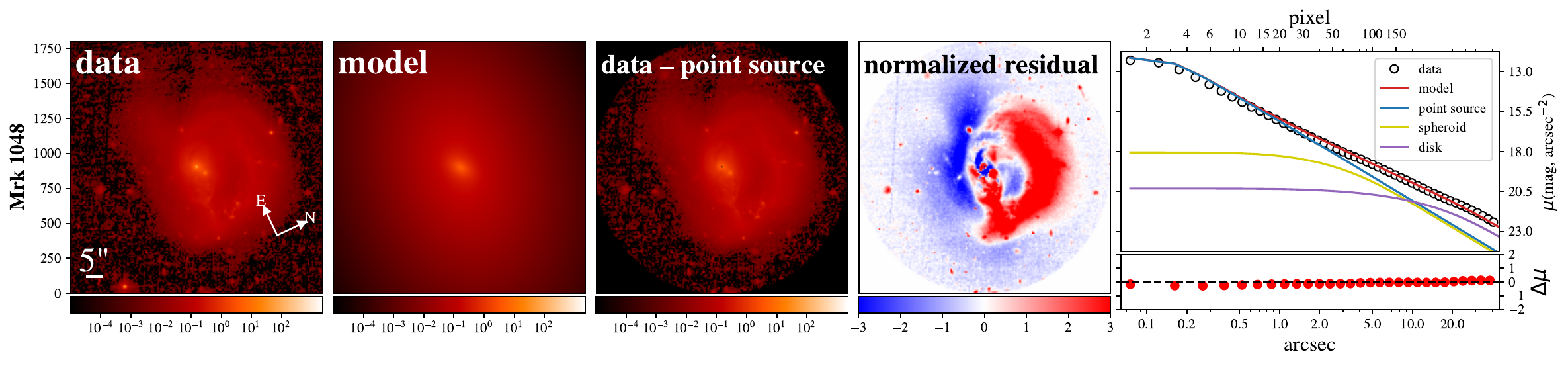}
   \includegraphics[width=\textwidth,height=0.16\textheight,keepaspectratio]{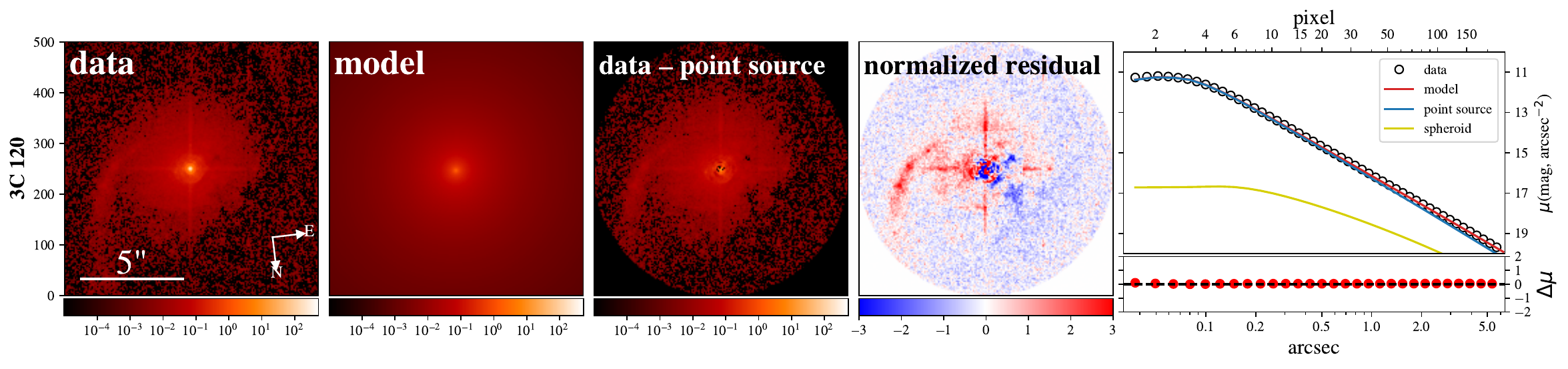}
   \includegraphics[width=\textwidth,height=0.16\textheight,keepaspectratio]{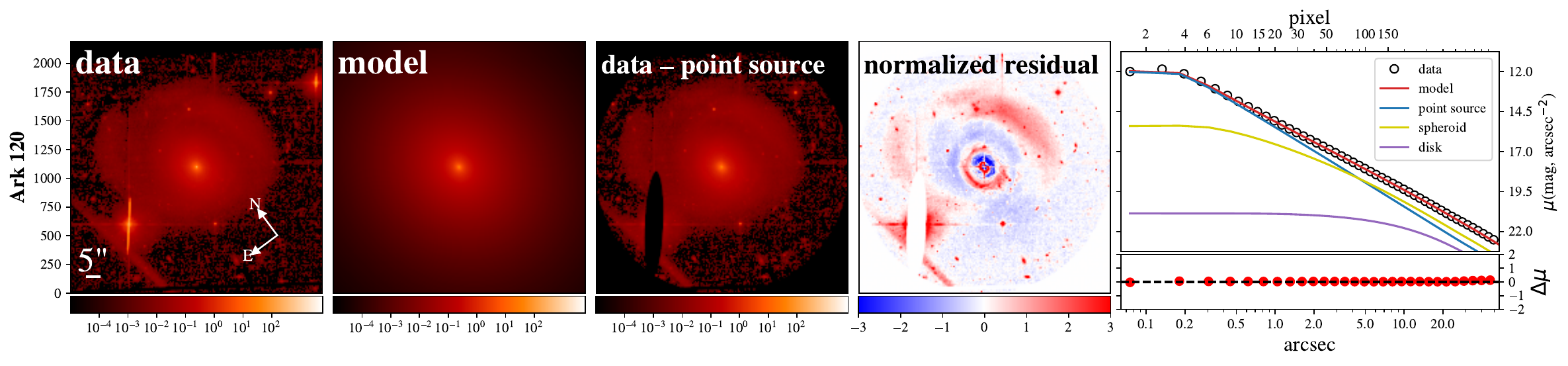}
   \includegraphics[width=\textwidth,height=0.16\textheight,keepaspectratio]{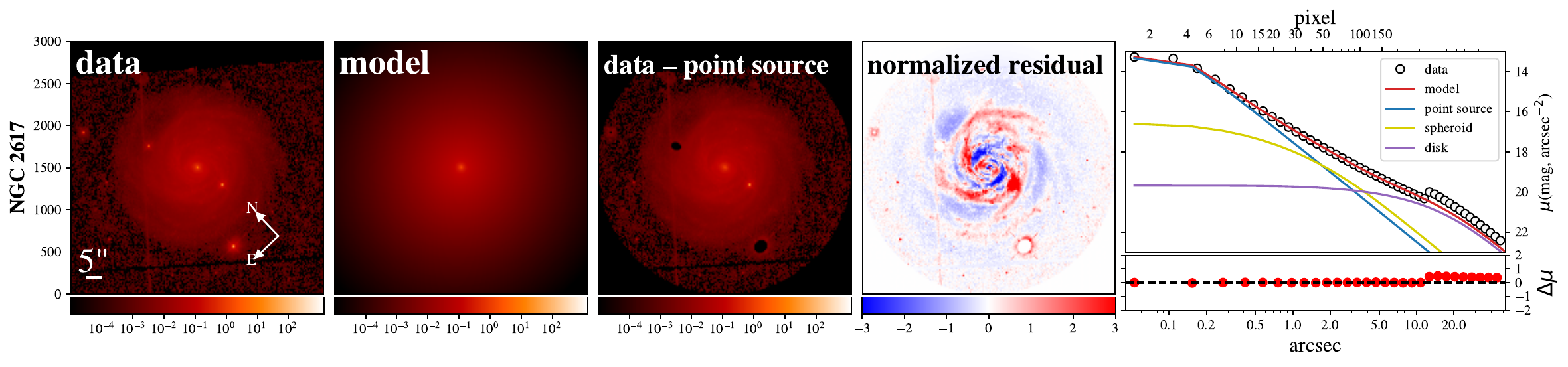}
  \includegraphics[width=\textwidth,height=0.16\textheight,keepaspectratio]{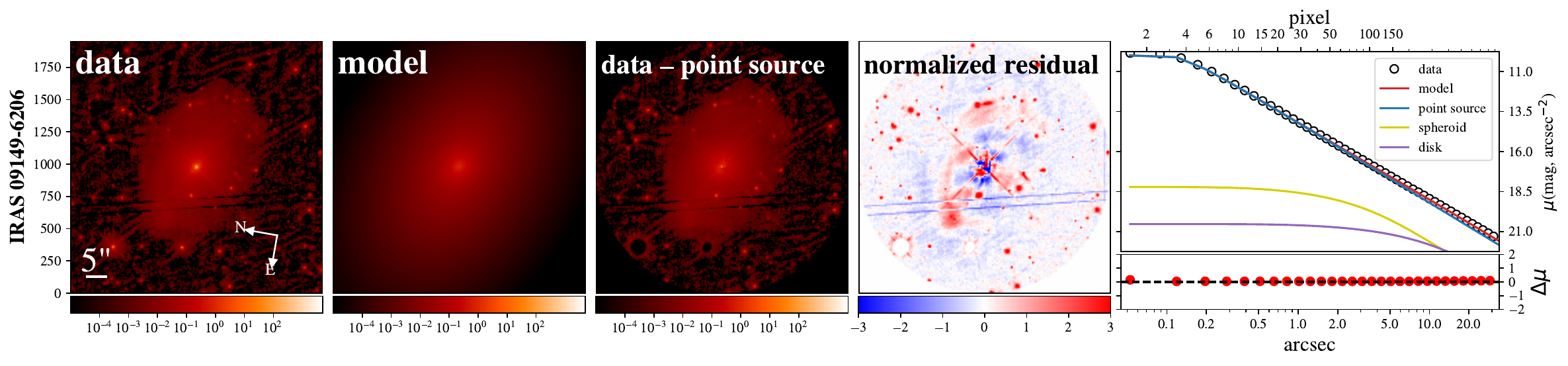}
   \caption{
Figure~\ref{fig:lenstronomy1} continued.
 }
              \label{fig:lenstronomy2}
\end{figure*}

\begin{figure*}[p]
  \centering
   \includegraphics[width=\textwidth,height=0.16\textheight,keepaspectratio]{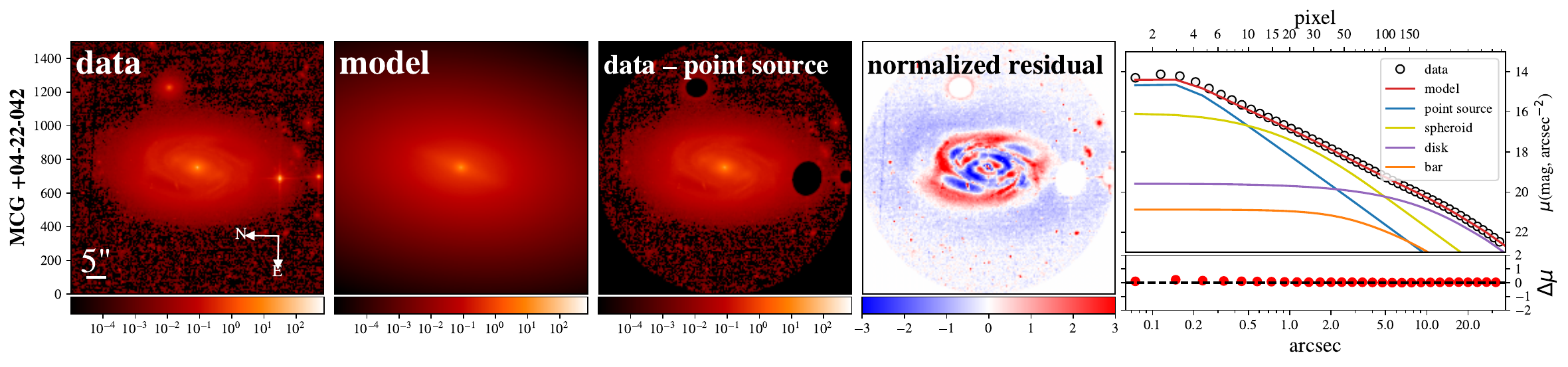}
   \includegraphics[width=\textwidth,height=0.16\textheight,keepaspectratio]{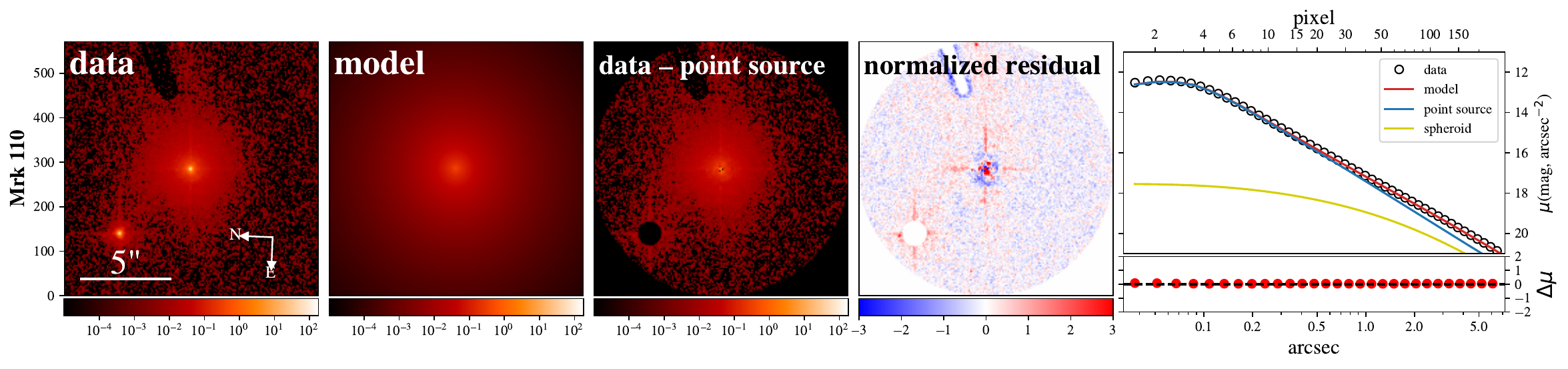}
   \includegraphics[width=\textwidth,height=0.16\textheight,keepaspectratio]{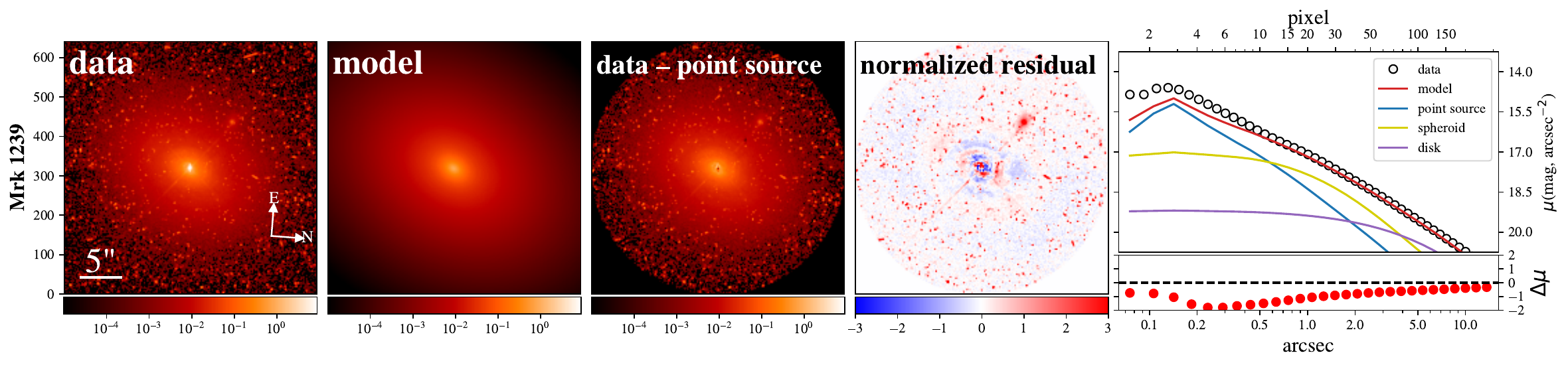}
   \includegraphics[width=\textwidth,height=0.16\textheight,keepaspectratio]{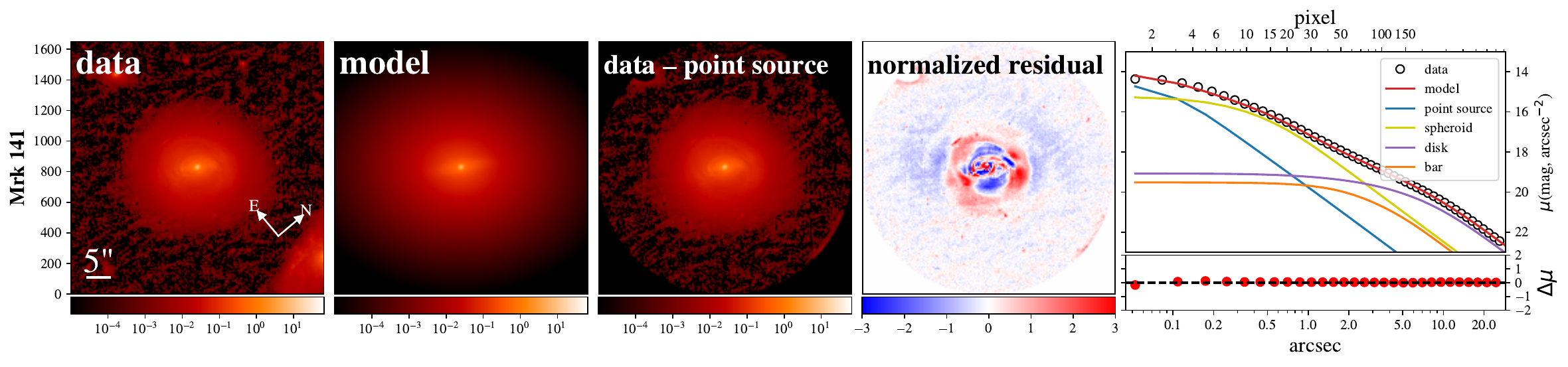}
  \includegraphics[width=\textwidth,height=0.16\textheight,keepaspectratio]{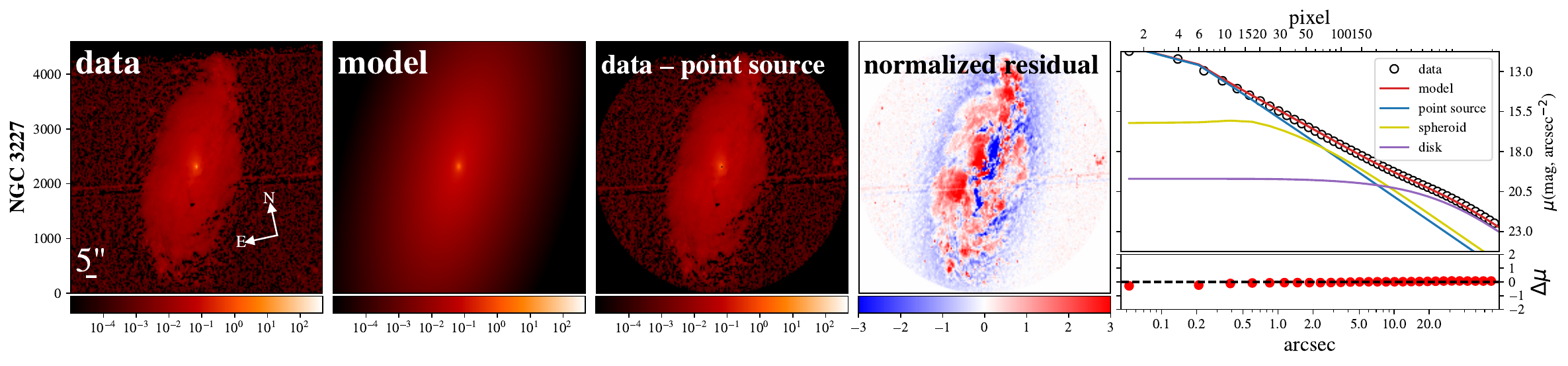}
   \caption{
Figure~\ref{fig:lenstronomy1} continued.
 }
              \label{fig:lenstronomy3}
\end{figure*}

\begin{figure*}[p]
  \centering
   \includegraphics[width=\textwidth,height=0.16\textheight,keepaspectratio]{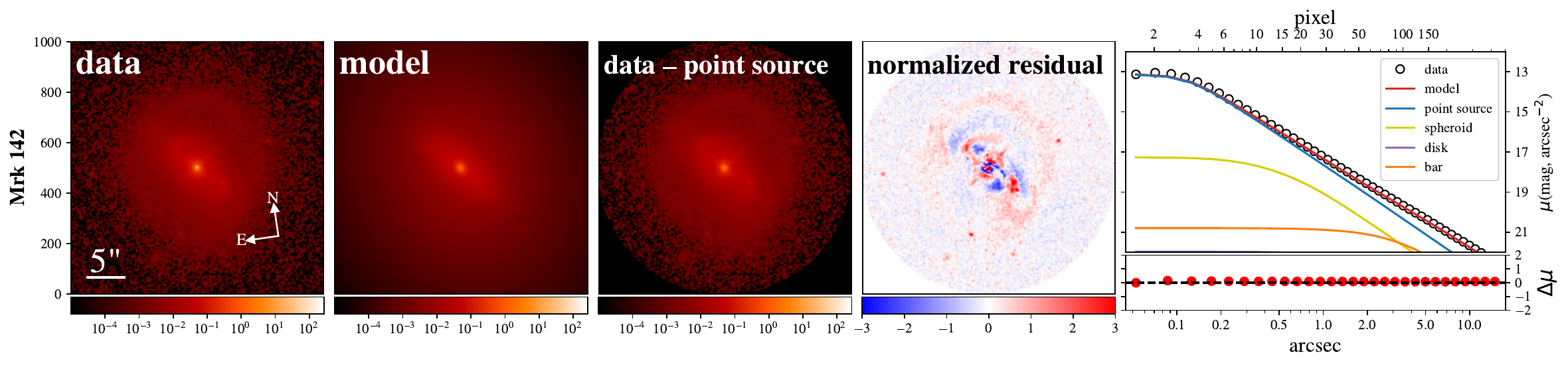}
   \includegraphics[width=\textwidth,height=0.16\textheight,keepaspectratio]{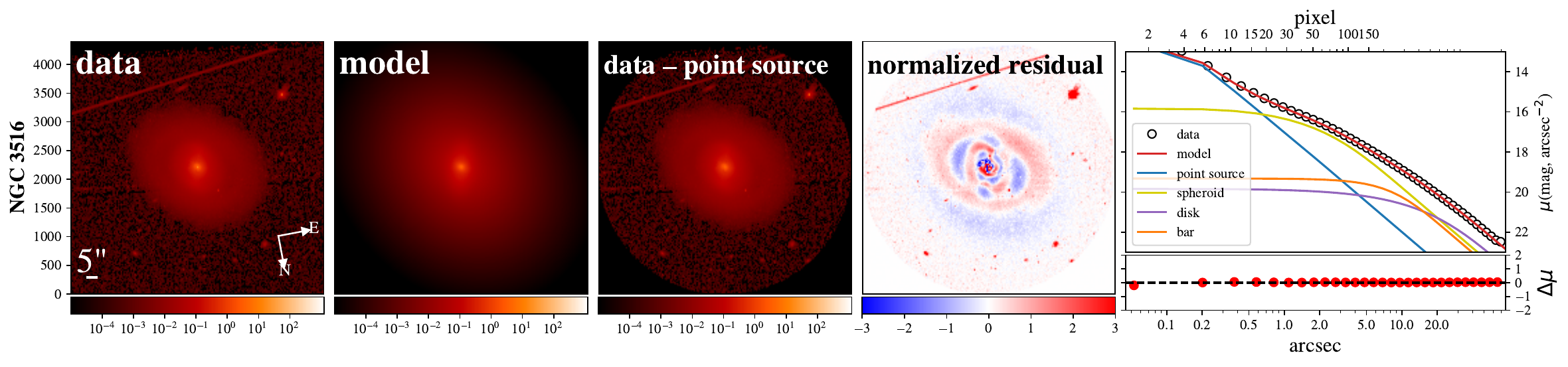}
   \includegraphics[width=\textwidth,height=0.16\textheight,keepaspectratio]{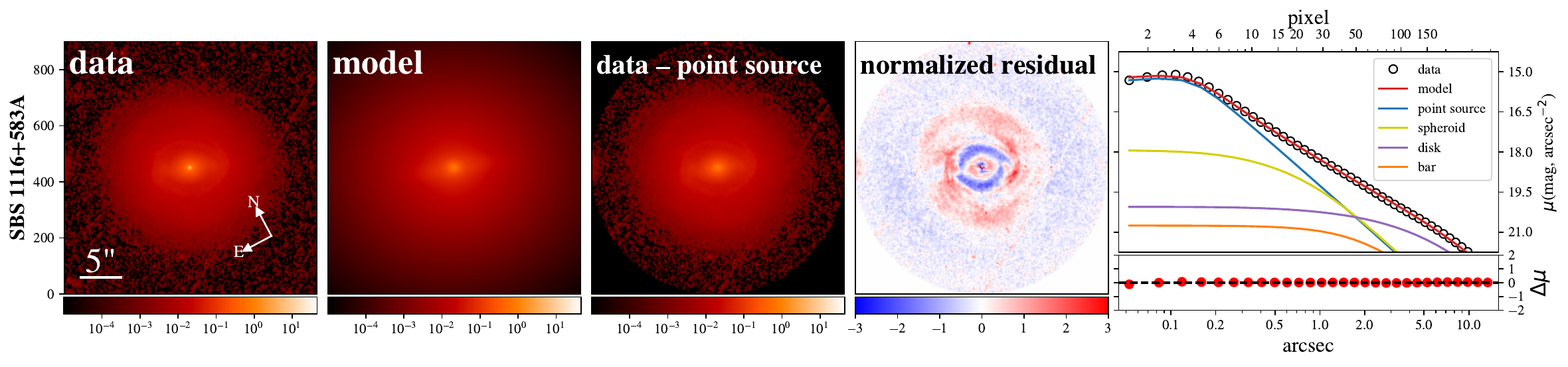}
  \includegraphics[width=\textwidth,height=0.16\textheight,keepaspectratio]{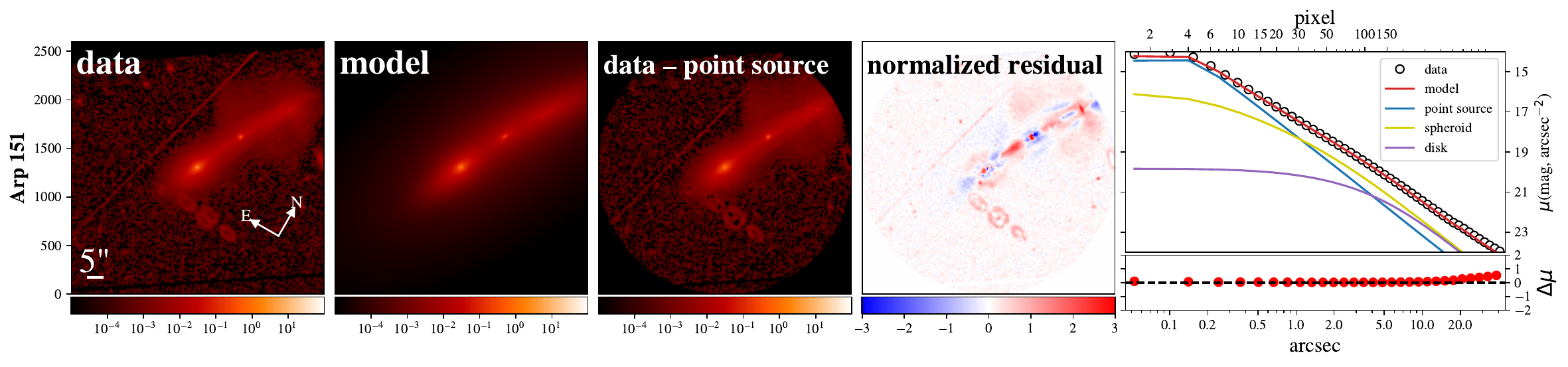}
   \includegraphics[width=\textwidth,height=0.16\textheight,keepaspectratio]{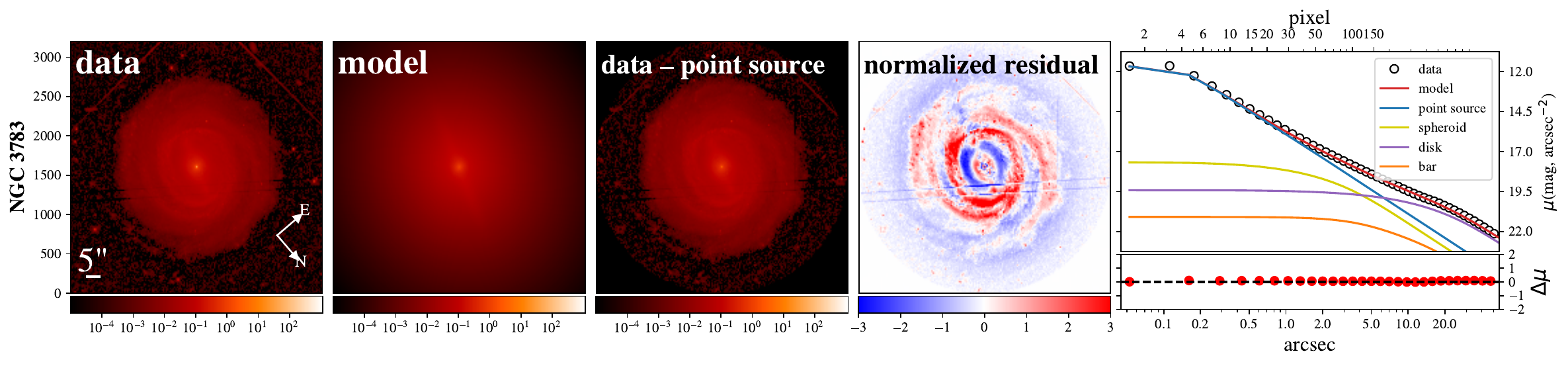}
   \caption{
Figure~\ref{fig:lenstronomy1} continued.
 }
              \label{fig:lenstronomy4}
\end{figure*}

\begin{figure*}[p]
  \centering
   \includegraphics[width=\textwidth,height=0.16\textheight,keepaspectratio]{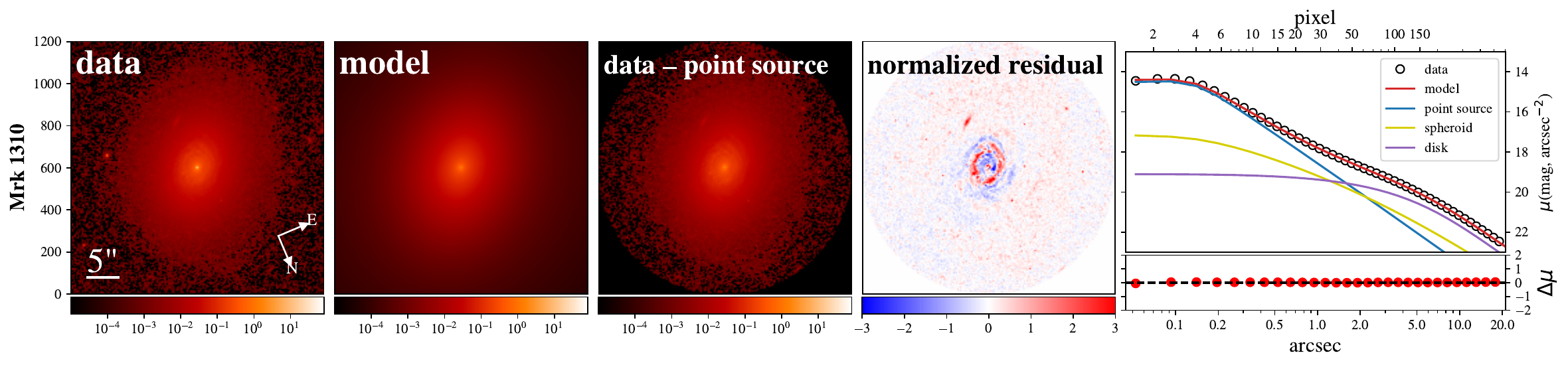}
   \includegraphics[width=\textwidth,height=0.16\textheight,keepaspectratio]{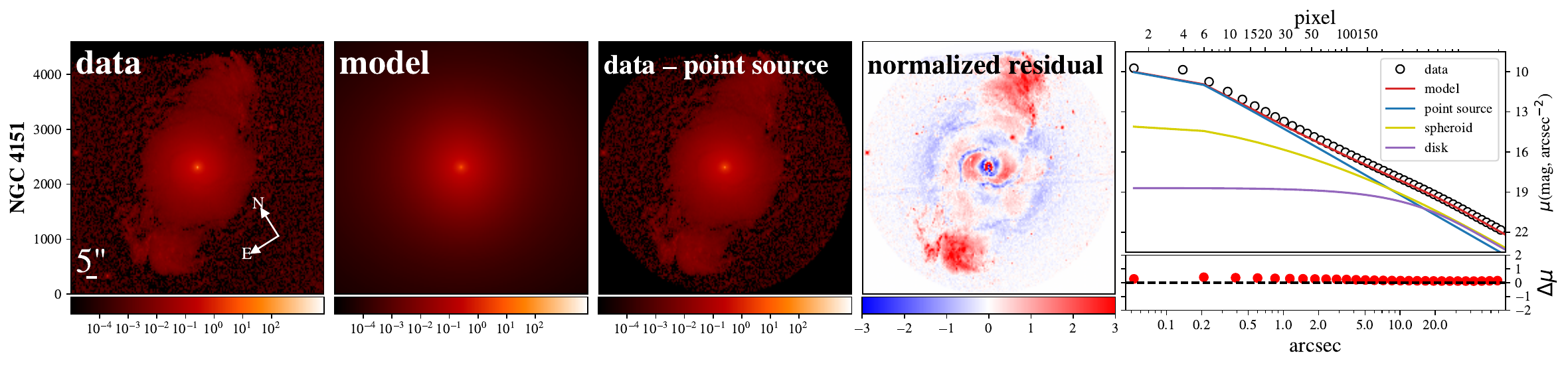}
   \includegraphics[width=\textwidth,height=0.16\textheight,keepaspectratio]{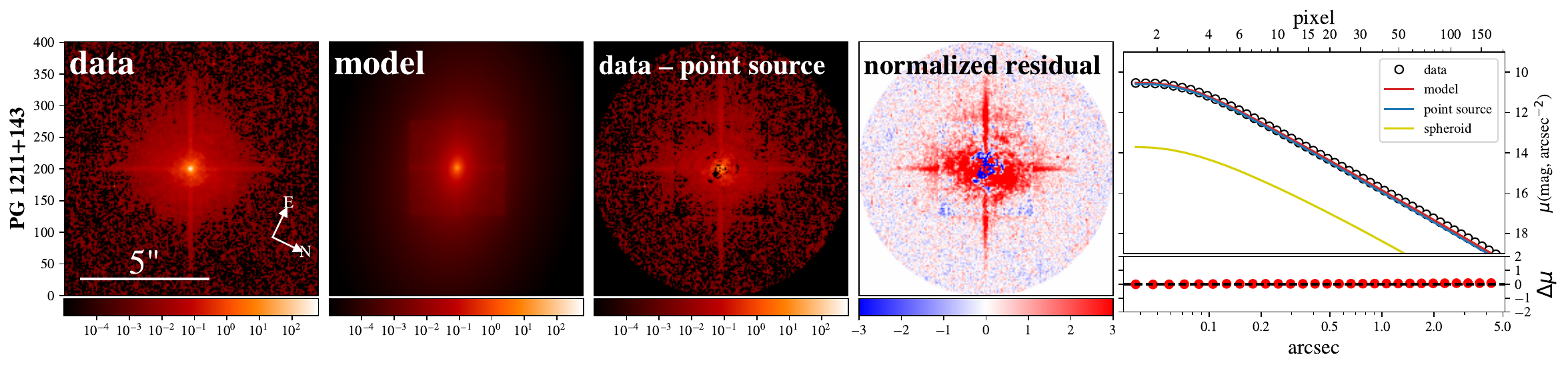}
   \includegraphics[width=\textwidth,height=0.16\textheight,keepaspectratio]{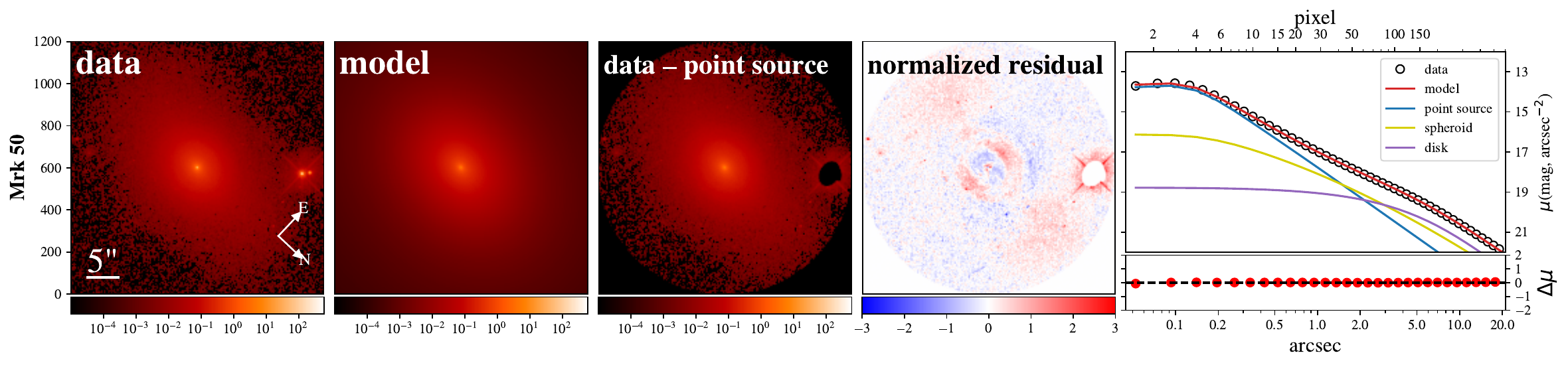}
    \includegraphics[width=\textwidth,height=0.16\textheight,keepaspectratio]{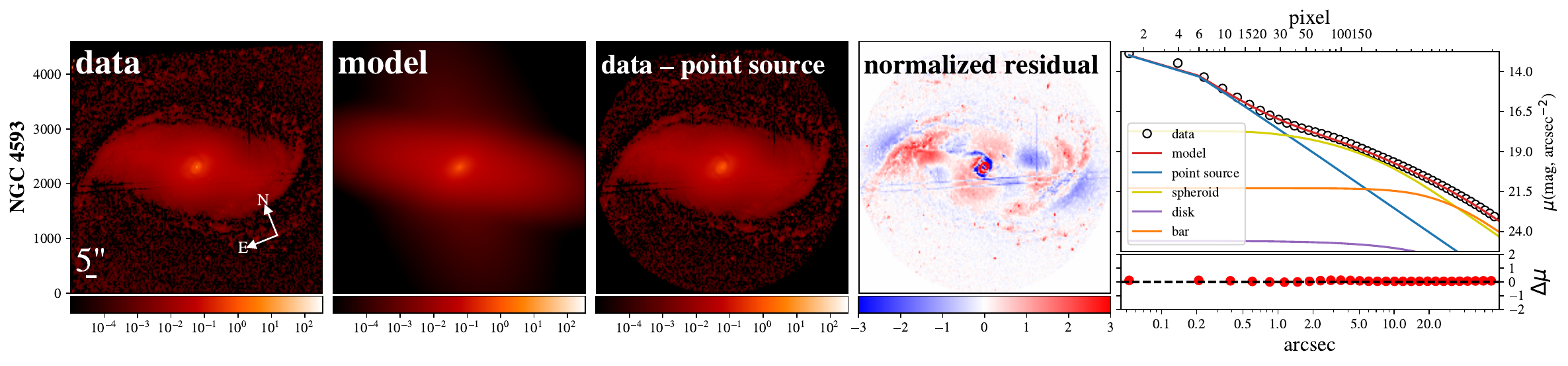}
   \caption{
Figure~\ref{fig:lenstronomy1} continued.
 }
              \label{fig:lenstronomy5}
\end{figure*}

\begin{figure*}[p]
  \centering
   \includegraphics[width=\textwidth,height=0.16\textheight,keepaspectratio]{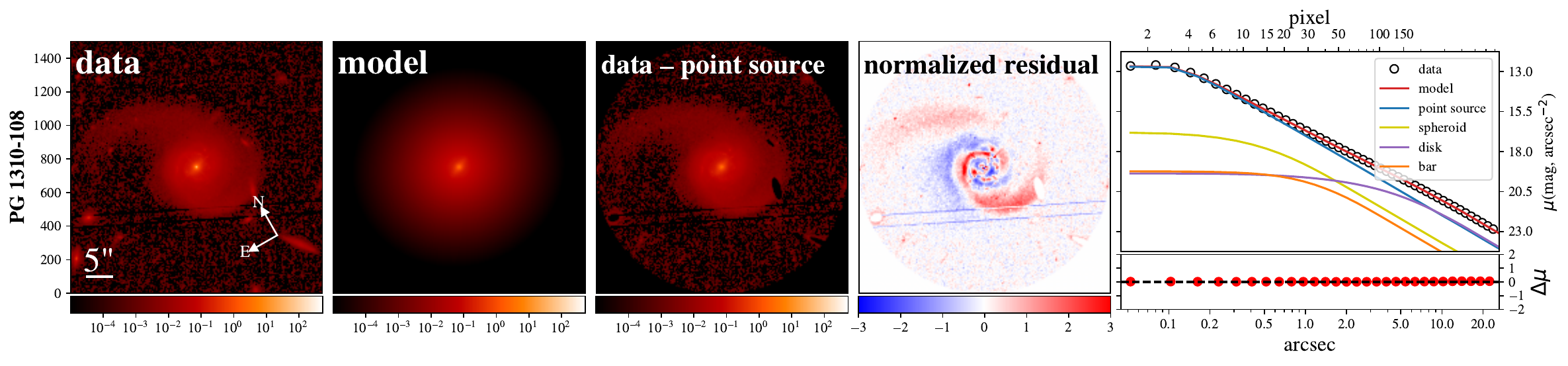}
   \includegraphics[width=\textwidth,height=0.16\textheight,keepaspectratio]{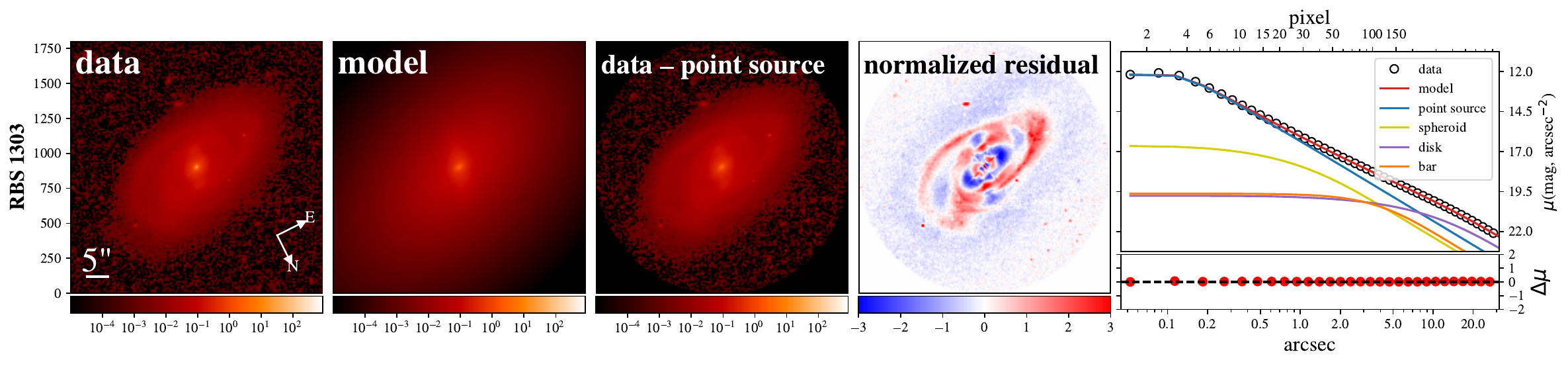}
   \includegraphics[width=\textwidth,height=0.16\textheight,keepaspectratio]{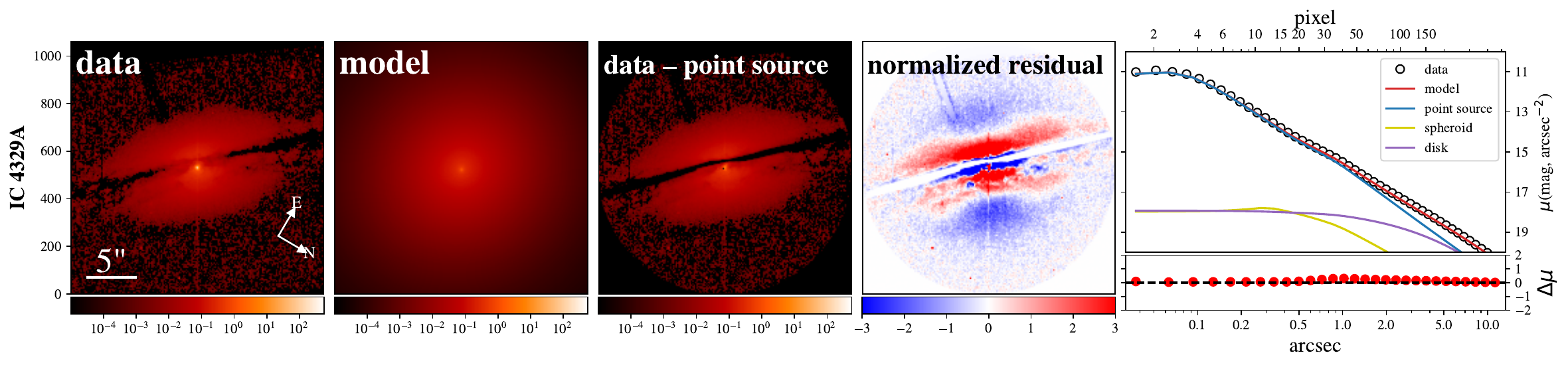}
  \includegraphics[width=\textwidth,height=0.16\textheight,keepaspectratio]{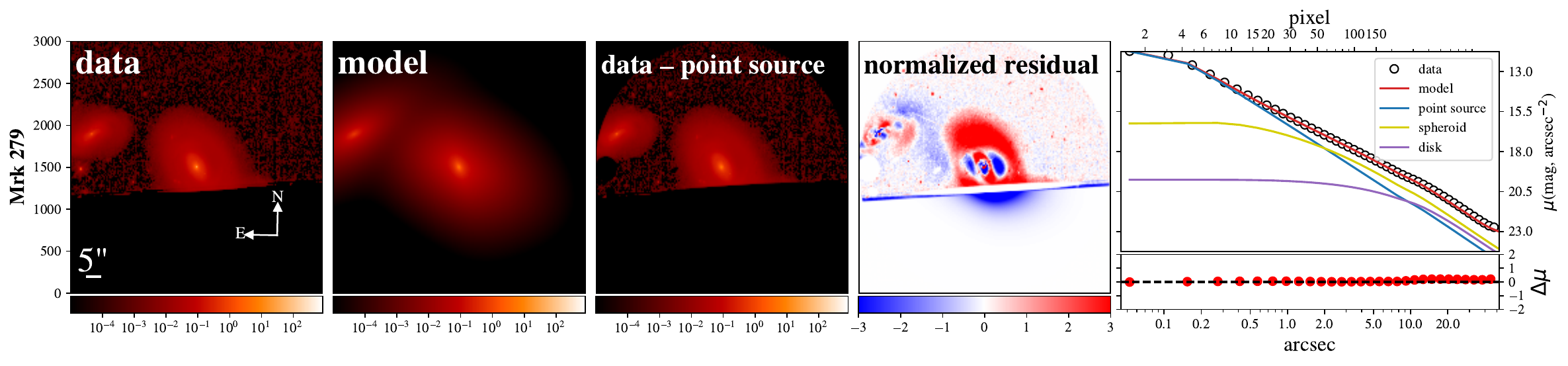}
   \includegraphics[width=\textwidth,height=0.16\textheight,keepaspectratio]{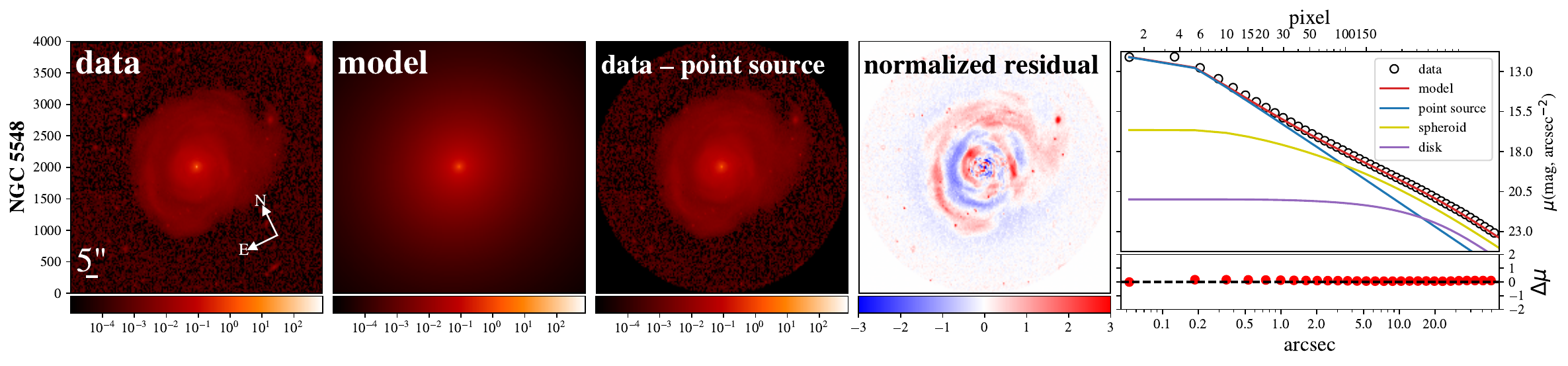}
   \caption{
Figure~\ref{fig:lenstronomy1} continued.
 }
              \label{fig:lenstronomy6}
\end{figure*}

\begin{figure*}[p]
  \centering
   \includegraphics[width=\textwidth,height=0.16\textheight,keepaspectratio]{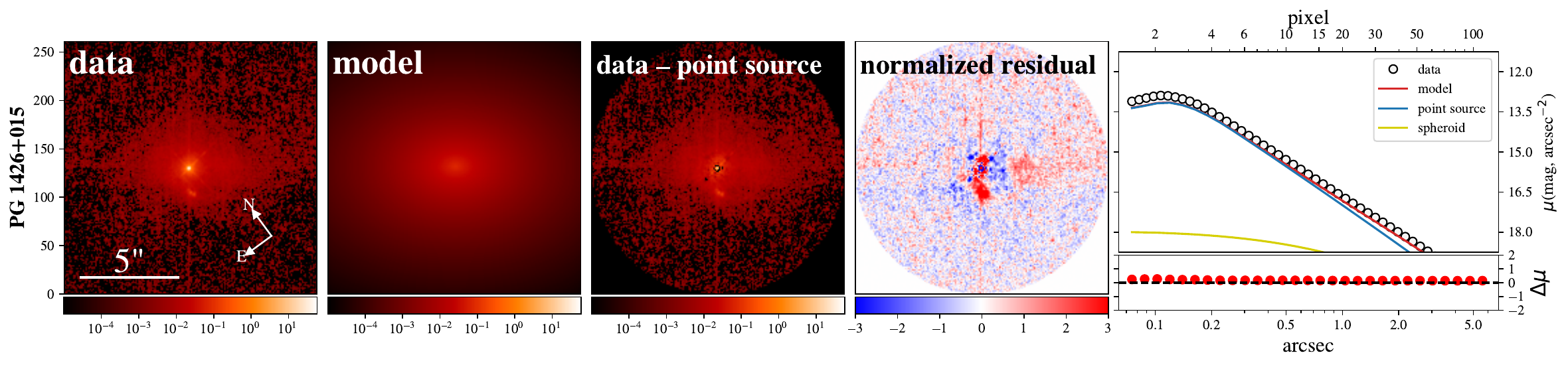}
   \includegraphics[width=\textwidth,height=0.16\textheight,keepaspectratio]{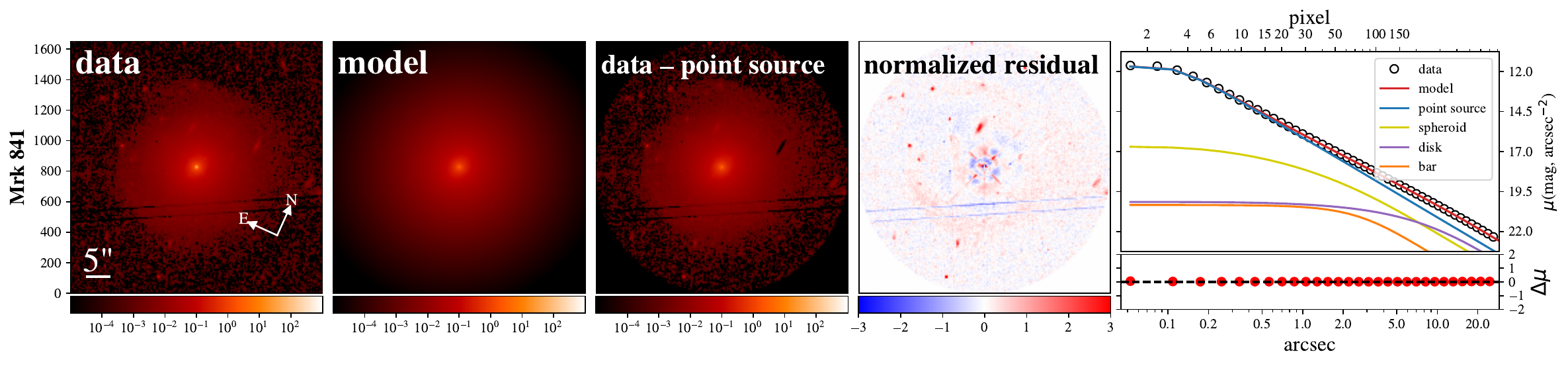}
      \includegraphics[width=\textwidth,height=0.16\textheight,keepaspectratio]{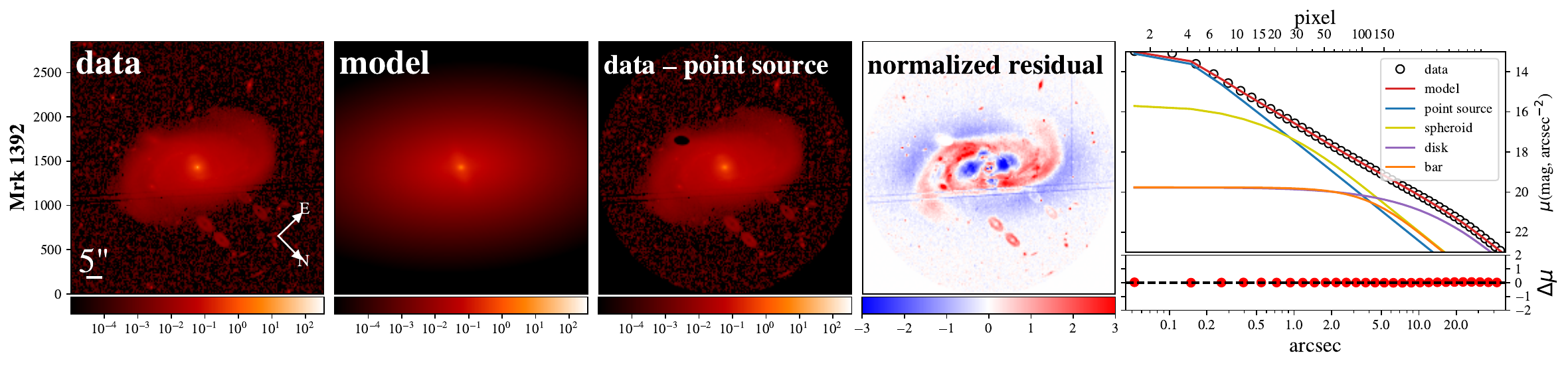}
   \includegraphics[width=\textwidth,height=0.16\textheight,keepaspectratio]{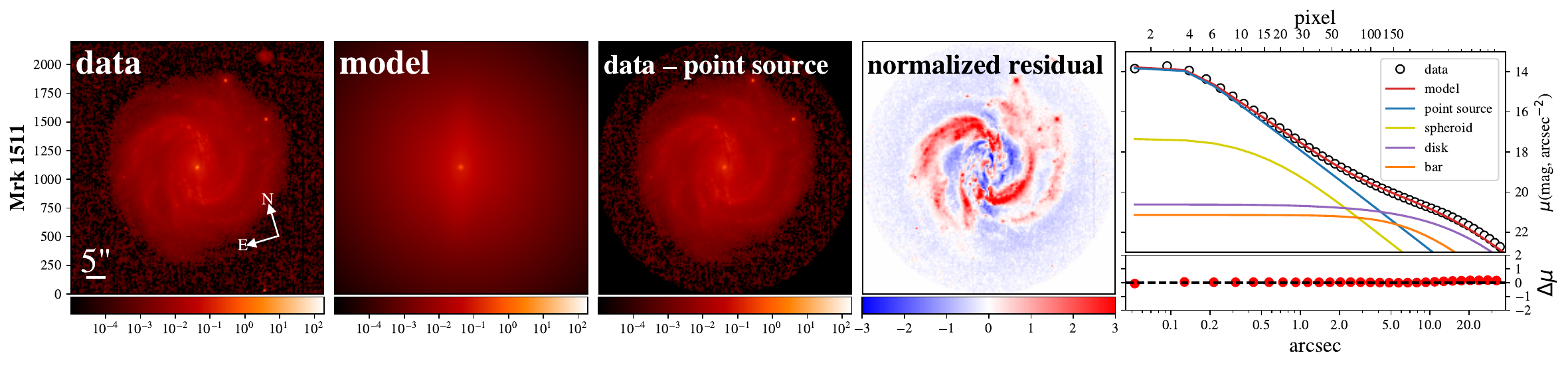}
   \includegraphics[width=\textwidth,height=0.16\textheight,keepaspectratio]{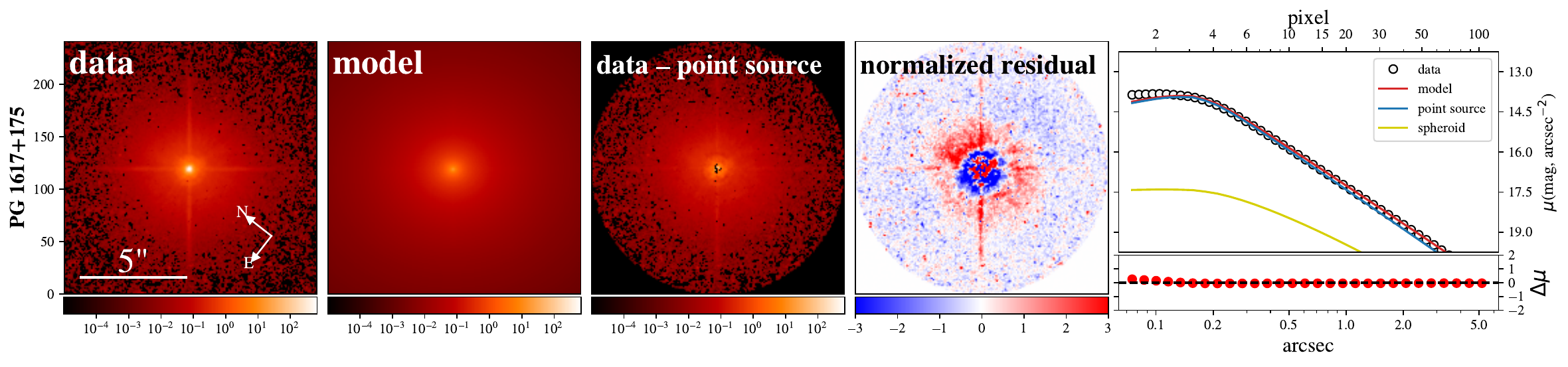}
   \caption{
Figure~\ref{fig:lenstronomy1} continued.
 }
              \label{fig:lenstronomy7}
\end{figure*}
   
\begin{figure*}[p]
     \centering
   \includegraphics[width=\textwidth,height=0.16\textheight,keepaspectratio]{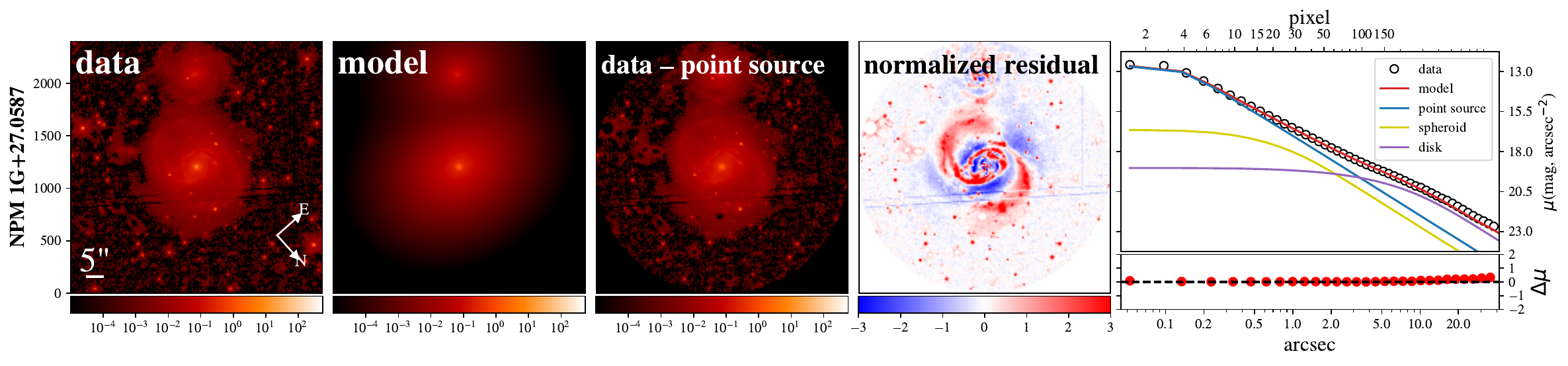}
     \includegraphics[width=\textwidth,height=0.16\textheight,keepaspectratio]{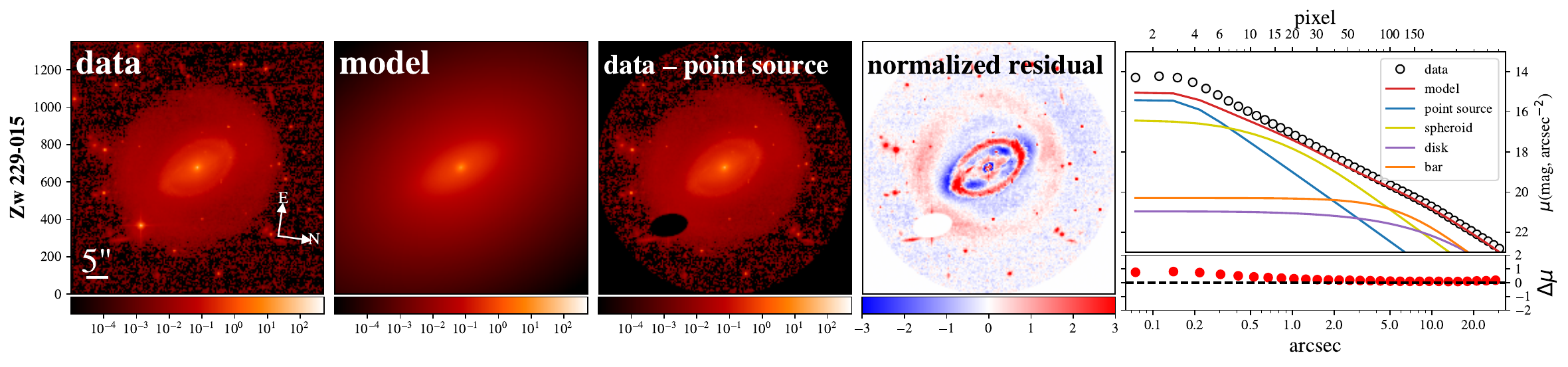}
   \includegraphics[width=\textwidth,height=0.16\textheight,keepaspectratio]{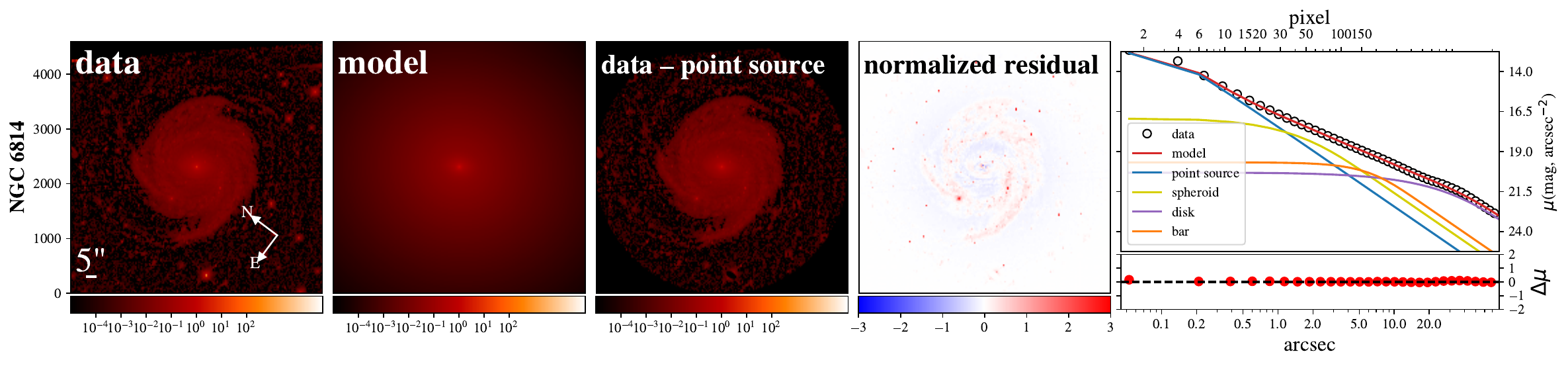}
   \includegraphics[width=\textwidth,height=0.16\textheight,keepaspectratio]{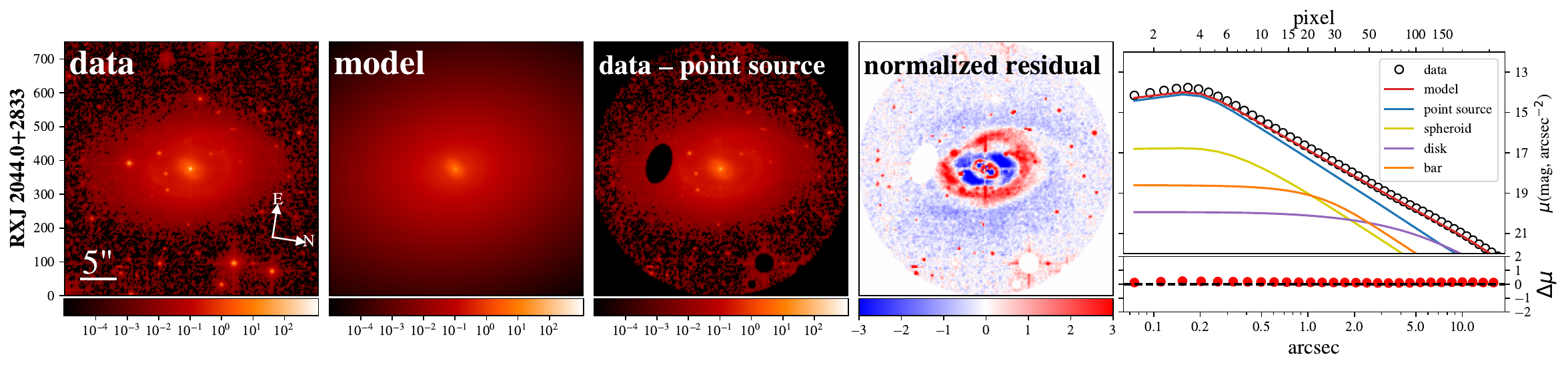}
   \includegraphics[width=\textwidth,height=0.16\textheight,keepaspectratio]{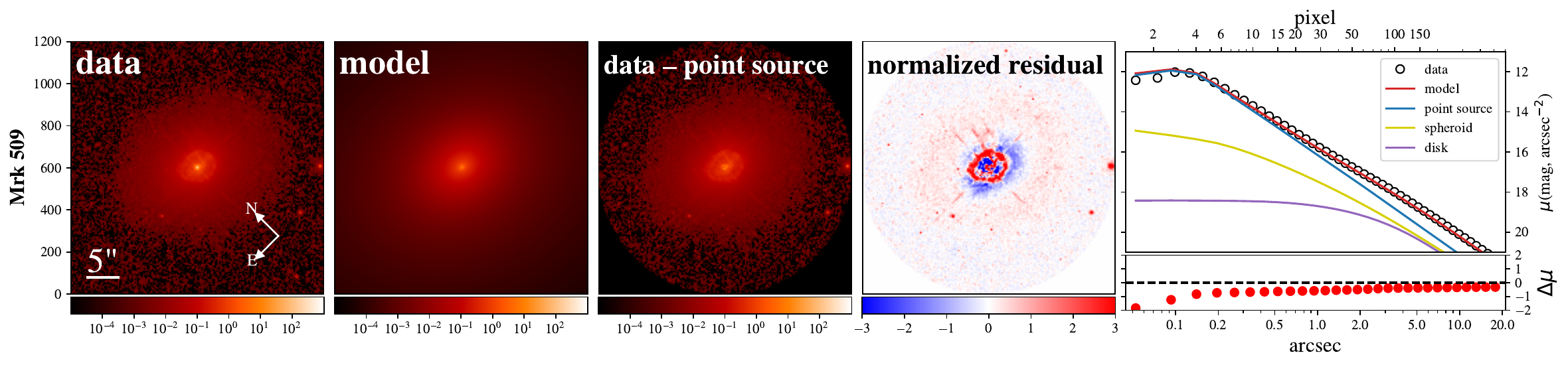}
   \caption{
Figure~\ref{fig:lenstronomy1} continued.
 }
              \label{fig:lenstronomy8}
\end{figure*}
   
\begin{figure*}[p]
     \centering
   \includegraphics[width=\textwidth,height=0.16\textheight,keepaspectratio]{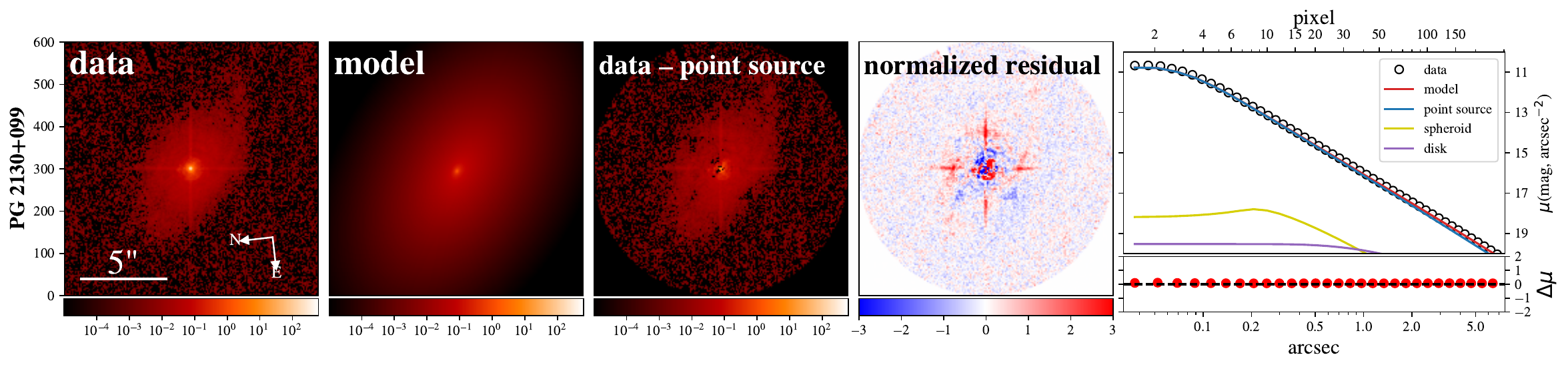}
    \includegraphics[width=\textwidth,height=0.16\textheight,keepaspectratio]{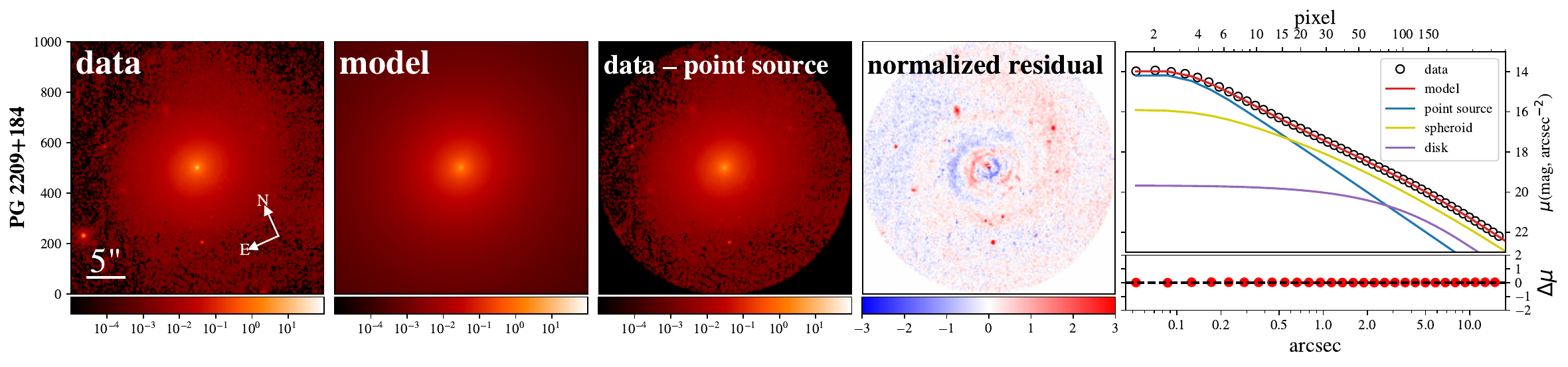}
   \includegraphics[width=\textwidth,height=0.16\textheight,keepaspectratio]{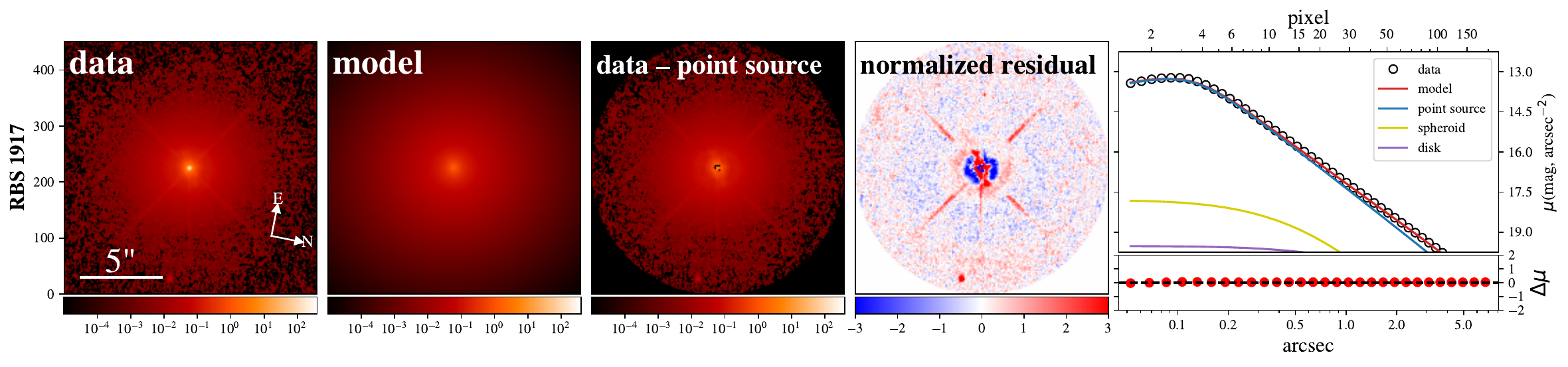}
   \includegraphics[width=\textwidth,height=0.16\textheight,keepaspectratio]{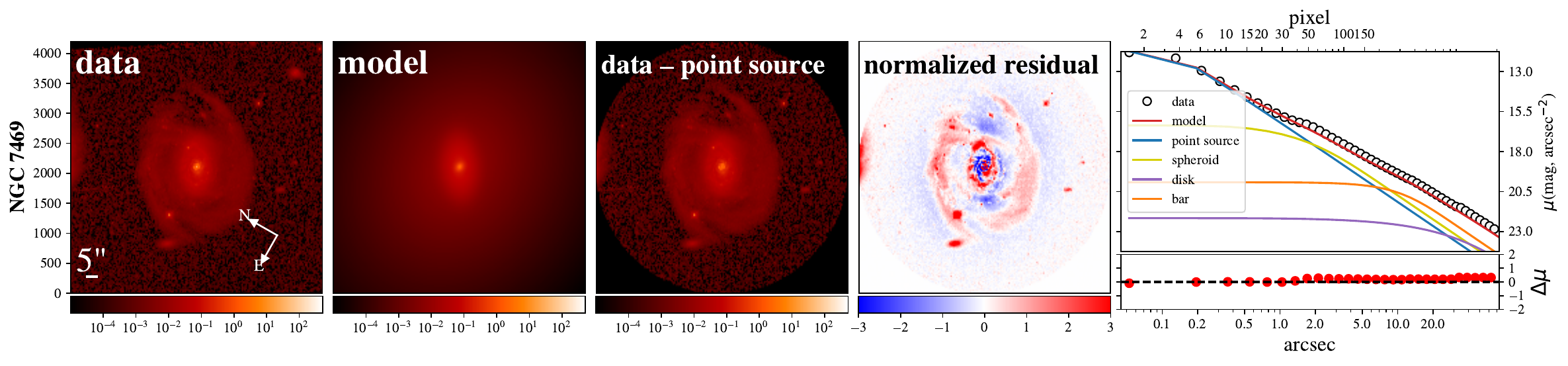}
   \caption{
Figure~\ref{fig:lenstronomy1} continued.
 }
              \label{fig:lenstronomy9}
\end{figure*}

\end{document}